\documentclass{aa501}
\usepackage{epsfig}
\begin{document}



   \title{Restoring the full velocity field in the gaseous disk of
the spiral galaxy NGC~157. }
\titlerunning{Gas velocity field in NGC~157}
   \subtitle{}

   \author{A.M. Fridman
          \inst{1,2}
          \and
          O.V. Khoruzhii
          \inst{1,3}
          \and
	  V.V. Lyakhovich
          \inst{1}
          \and
	  O.K. Sil'chenko
          \inst{2}
          \and
	  A.V. Zasov
          \inst{2}
          \and
          V.L. Afanasiev
          \inst{4}
          \and
	  S.N. Dodonov
          \inst{4}
          \and
	  J. Boulesteix
          \inst{5}
          }

   \offprints{A.M. Fridman}

   \institute{Institute of Astronomy of the Russian Academy of Science,
48, Pyatnitskaya St., Moscow, 109017, Russia\\
              email: afridman@inasan.rssi.ru
         \and
Sternberg Astronomical Institute,
Moscow State University, University prospect, 13,
Moscow, 119899, Russia
         \and
National Research Center "Troitsk Institute for Innovation and
Thermonuclear Researches", Troitsk,
Moscow reg., 142092, Russia
         \and
Special Astrophysical Observatory of the Russian
Academy of Sciences, Zelenchukskaya, 377140, Russia
	\and
	Observatoire de Marseille, Place le Verrier,
F-13248, Marseille, Cedex 04, France
             }

   \date{Received ; accepted }

\abstract{
This paper is the first in a series of articles devoted to
the construction and analysis of three-component vector velocity
fields in the gaseous disks of spiral galaxies, and to the discovery
of giant anticyclones near corotation which were predicted earlier. We
analyse the line-of-sight velocity field of ionized gas in the spiral
galaxy NGC~157 which has been obtained in the H$\alpha$ emission line at
the 6m telescope of SAO RAS. The field contains more than 11,000 velocity
estimates. The existence of systematic deviations of the
observed gas velocities from pure circular motion is shown. A detailed
investigation of these deviations is undertaken by applying a
recently formulated method based on Fourier analysis of the azimuthal
distributions of the line-of-sight velocities at different distances
from the galactic center.  To restore the three-component
vector velocity field from the observed line-of-sight velocity
field, two assumptions were made: ~1) the perturbed surface density and
residual velocity components can be approximated to the formulae
$C_i\,\cos(2\varphi+F_i)$, where $C_i$ and $F_i$ are respectively
the amplitude and the phase; ~2) the perturbed surface
density and residual velocity components satisfy the Euler equations.
The correctness of both assumptions is proven on the basis of the
observational data. As a result of the analysis, all the main parameters
of the wave spiral pattern are determined:  the corotation radius, the
amplitudes and phases of the gas velocity perturbations at different
radii, as well as the velocity of circular rotation of the disk corrected
for the influence of the velocity perturbations connected with the
spiral arms.  Finally, a restoration of the vector velocity field in
the gaseous disk of NGC~157 is performed. At a high confidence
level, the presence of the two giant anticyclones in the reference
frame rotating with the spiral pattern is shown; their sizes and the
localization of their centers are consistent with the results of the
analytic theory and of numerical simulations.  Besides the
anticyclones, the existence of cyclones in any residual velocity field is
predicted. In the reference frame of the spiral pattern, detection of
such cyclones is possible in the galaxies with a radial gradient
of azimuthal residual velocity steeper than that of the rotation
velocity.
}

\maketitle

\section{Introduction}

The present work is devoted to the analysis of line-of-sight
velocities of the gas in the spiral galaxy NGC~157. We attempt to evaluate
all three vector components of the perturbed gas velocity in the spiral
density wave to find the angular velocity of the spiral pattern  and
the resonance locations, to obtain the rotation curve corrected for
non-circular motions connected with the wave structures, and also to
demonstrate the existence of a vortex structure.

Evidently, any method of restoration of the three-component vector velocity
field from the observed one-component line-of-sight velocity field needs
some additional assumptions, and the correctness of the restoration
completely depends on the adequacy of those assumptions. Here two
assumptions were made: 1)  the perturbed surface density and residual
velocity components in the NGC~157 can be approximated to the form
$C_i\,\cos(2\varphi+F_i)$, where $C_i$ and $F_i$ are respectively
correspondingly the amplitude and the phase; 2) the perturbed surface
density and residual velocity components satisfy the Euler equations. In
the paper we show that both assumptions agree with the observational data.
First, the predominance of the second Fourier harmonic of the brightness
over others allows to adequately present the perturbed surface density in
the form $\tilde \sigma$ $=$ $C_\sigma(r,\,t)\,\cos[2\varphi -
F_\sigma(r)]$, whereas the prevalence of the second, third and the sine
component of the first Fourier harmonics in the residual velocity field
means that the perturbed velocity components were chosen correctly, in
a similar manner. Second, analysis of the radial dependencies of the
harmonic phases gives evidence for the existence of a connection
between the line-of-sight velocity disturbances of the gas and the
observed spiral arms, in agreement with what follows from
the Euler equations.

In Sect.~2, a description of the observations of the galaxy is
given. In Sect.~3, the results of an  analysis of the line-of-sight
velocity field under the approximation of pure circular motion are
presented; we compare these results with the earlier published
observational data and discuss residual line-of-sight velocities obtained
by subtracting the model velocity field from the observed one. We find
that the deviations from pure circular motion demonstrate a systematic
behaviour. In Sect.~4, a simple model of the line-of-sight velocities
is described, which takes into account motions in the two-armed
density wave.  It is shown that in this case the first three Fourier
harmonics dominate throughout the line-of-sight velocity field. In
Sect.~5, we present the results of the harmonic analysis of the observed
line-of-sight velocity field which prove that the perturbed surface
density and residual velocity components satisfy the Euler equations (i.e.
the proof of the wave nature of the spiral structure). After that,
restoration and analysis of the vector velocity field of the gas is
performed. In Sect.~6, we show that the Fourier analysis of the
line-of-sight velocity field allows us to restore the velocity field
in the plane of the galaxy with a high degree of confidence. In
Sect.~7, the corotation position is determined directly from the
observations. It is shown that the velocity field in the reference frame
rotating with the spiral pattern clearly demonstrates the presence of two
banana-like anticyclones, the centers of which lie near the corotation.
The last Section contains the main conclusions.

\section{Observations and primary data reduction}

\begin{figure}
\epsfig{figure=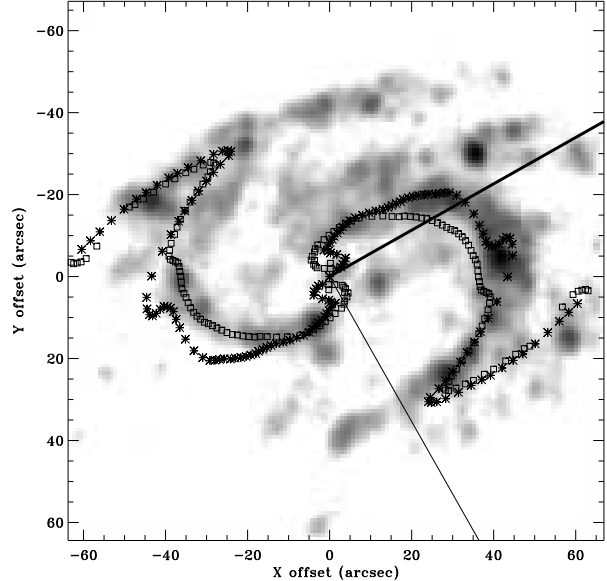,width=8.0cm,
bbllx=75pt, bblly=190pt, bburx=535pt, bbury=645pt, clip=}
\caption[]{A monochromatic image of NGC~157 in the H$\alpha$ emission
line. North is roughly to the left (PA$_y$ $=$ $-76^\circ$).
The phase curve of the second Fourier harmonic of the H$\alpha$
brightness map is superimposed. Asterisks show the azimuth positions of
the maxima of the second harmonic at each radius.  Positions of the
maxima of the second Fourier harmonic of the NGC~157 $K$-band
brightness map (kindly provided by S.D. Ryder) are also shown by squares.
}
\label{f-2}
\end{figure}

NGC~157 (Fig.~\ref{f-2}) is a rather bright  ($B_T$ = 11.00) and close
($D\sim 21$ Mpc for $H_0$ = 75 km s$^{-1}$ Mpc$^{-1}$)  isolated galaxy of
SAB(rs)bc type (de Vaucouleurs et al. 1991). Its well--developed
Grand--Design spiral structure is classified as Arm class 12 by Elmegreen
\& Elmegreen (1984).  This galaxy is characterised by mildly enhanced star
formation, which follows from its rather high brightness in the
H$\alpha$ line and from far infrared flux densities (Ryder et al.
1998). The large angular size (the isophotal radius $D_o / 2$ = $2.'13$,
or $\sim$ 13 kpc) and the moderate disk inclination (axis ratio $b/a$ =
0.65) make this galaxy very convenient for kinematic and photometric
measurements.

The optical morphology of this galaxy is rather complex.  A
blue image (Lynds 1974) shows a dusty and more or less symmetrical disk
with two principal spiral arms and without any hint of a  bar.
Nevertheless there is some evidence of a weak bar, lacking bright
HII regions, in the inner part of the disk, noticeable in the red
($R$-band) image of the galaxy (Sempere \& Rozas 1997). Note, however,
that this bar does not reveal itself in the kinematics of H{\sc i} (Ryder et
al. 1998). The spiral structure of NGC 157 is rather regular up to about
$1'$, or $\simeq$ 6.1 kpc, from the center, and, by sight, becomes
flocculent or multiarmed at the periphery. As follows from the H$\alpha$
image of the galaxy, the two principal arms are not symmetrical with
respect to the center (Sempere \& Rozas 1997).  According to Elmegreen et
al. (1992) such an asymmetry may be responsible for the additional
m=1 component of the spiral wave structure, driven by the two-armed spiral
in the inner disk.  Note also that one of the two predominant spiral arms
is brighter in H$\alpha$ than the other.

Radio observations in the H{\sc i} line and at the 1.4 GHz continuum
(Ryder et al. 1998) show that there is a close, though not strict
correspondence of radio brightness and H$\alpha$ surface
brightness. In addition, NGC~157 appears to possess an extended
warped H{\sc i} disk, which stretches out well beyond the optical borders.
The rotation velocity of the H{\sc i} disk amounts to about half the
maximal velocity of the optical disk, which enables us to conclude that
the galaxy has (1) a falling rotation curve and  (2) an extended
low-density halo beyond $r\sim 8-10$ kpc .  The latter contains a small
fraction of the total mass (about 10\%) within the isophotal radius, so
its gravitational input in the region of the regular spiral structure is
quite negligible.

Rather bright H$\alpha$ emission is observed not only inside
the spiral arms, but also between them. This circumstance is favourable
for the kinematic study of the ionized gas. The first investigation
of the disk kinematics was carried out by Burbidge et al. (1961) who
derived a  slowly rising rotation curve from two long-slit
spectra.  Later Zasov and Kyazumov (1981), using long-slit spectral
observations at different position angles, found a significant drop of the
rotational velocity at  $r \simeq$ 50$''$--55$''$ in the
north-eastern part of the disk.  Recent H{\sc i} observations (Ryder et al.
1998) have confirmed the reality of this peculiar motion and have given
evidence that it is caused both by a real declinion of the rotation
curve and by asymmetry in the line-of-sight velocity distribution along
the given positional angle. The optical rotation curve of the inner part
of this galaxy was also obtained by Afanasiev et al. (1988). It was found
that the rotation curve slowly rises, with $V_{\rm rot} \approx$ 100 km
s$^{-1}$ at $r = 1 - 2$ kpc and reaching a flat maximum of $\sim $ 200 km
s$^{-1}$ at $r \sim 5-6$ kpc.

The well-defined spiral structure in the inner angular radius of one
arcminute makes NGC 157 suitable for the detailed investigation of
density wave propagation in the disk. Analyzing the positions of different
morphological features in this galaxy, Elmegreen et al. (1992) found
that corotation takes place at about $r \sim 56\arcsec$, so that the
corotation circle limits the region of regular spiral structure.
Note, however, that the determination of the location of
resonances by morphological tracers involves a high degree of
uncertainty (see the discussion in Sempere \& Rozas 1997). The
alternative approach used by Sempere \& Rozas (1997) is based on
numerical simulations of the motions of interstellar molecular clouds in
the stellar disk potential, as calculated from the $R$-band image. The
resulting distribution of molecular clouds has then been compared
with the observed spiral structure of the galaxy. Although the results are
model dependent, the obtained pattern speed in the best fit model is
$\Omega_p$ =40 km s$^{-1}$ kpc$^{-1}$, which corresponds to a
corotation radius of 50\arcsec, close to the value claimed earlier by
Elmegreen et al. (1992).

Here we use a quite different method, the most direct one, to
investigate the interconnection between the spiral wave
and the kinematic behaviour of the interstellar gas. Our method is
based on a Fourier analysis of the observations of the line-of sight
velocity field (Lyakhovich et al. 1997, referred hereafter as L97). Some
preliminary results of this work were given in Fridman et al. (1997,
referred hereafter as F97).

\begin{figure}
\epsfig{figure=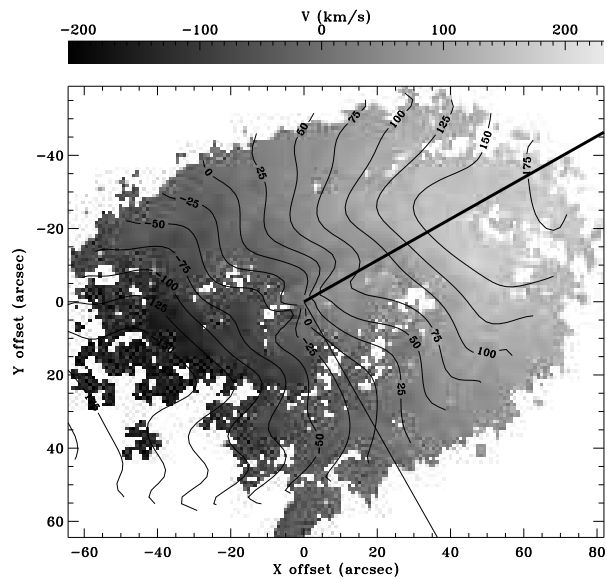,width=8.0cm,
bbllx=75pt, bblly=138pt, bburx=535pt, bbury=575pt, clip=}
\caption{ The line-of-sight velocity field of the gas in NGC~157 with
isovelocity contours overlaid. Systematic deviations from pure
circular motion, connected with the spiral arms, are clearly seen.}
\label{f-1}
\end{figure}

To obtain more detailed information about the kinematics of gaseous
clouds in NGC~157, we have undertaken observations in the H$\alpha$
emission line with the scanning Fabry-Perot Interferometer
(Dodonov et al. 1995) at the 6-m Telescope of the Special Astrophysical
Observatory of the Russian Academy of Sciences on October 24, 1995.  The
Interferometer was mounted in a focal reducer ($0\farcs46$ per pixel
scale) and installed at the prime focus of the telescope. To obtain 32
exposures 180 seconds each, we used the IPCS ($512 \times
512$~pixels) as a detector (Afanasiev et al. 1986, 1995).  Atmospheric
conditions were photometric with a seeing of 2.5~arcsec. The
Fabry-Perot Interferometer was tuned at the 501st order and gave a
velocity sampling of 19~km s$^{-1}$. The order-separating filter with
bandpass FWHM of 10 \AA~ ($\lambda_c$ $=$ 6613 \AA) was used to extract
the chosen etalon order. Phase calibration was made using a neon line
at $\lambda =$ 6598.95~\AA.

The data were reduced with the ADHOC software package,
developed in Marseille Observatory (Boulesteix 1993). The resulting data
cube of $256 \times 256 \times 32$ pixels was created with a
spatial sampling of 0.92~arcsec and a  velocity sampling of 19~km
s$^{-1}$. The spatial resolution of 2.5~arcsec corresponds to 250-300~pc,
at the distance of NGC~157 with $H_0$ $=$ 75~km s$^{-1}$ Mpc$^{-1}$.
The line-of-sight velocity field as reduced from the data cube is
shown in Fig.~\ref{f-1}. The isovelocity contours are superimposed on this
image.  Even on this figure one can clearly see the systematic
perturbation of the velocity field connected with the observed spiral
arms.  That becomes possible, in the first place, due to the
continuous character of the velocity measurements, filling both the spiral
arms and the interarm regions, which is what makes this
particular galaxy very convenient for a detailed analysis of the velocity
field.

\section{Pure circular motion model and residual velocities}

Systematic distortions of the line-of-sight velocity fields of a gas,
induced by galactic spiral structure, are well known from
observations (Lin et al. 1969; Yuan 1969; Visser 1980). The
amplitude of such distortions depends on a lot of physical and geometrical
parameters, particularly, on the wave amplitude, the form of the rotation
curve, and the corotation position. The parameters listed above are to
be determined.

The usual approach to the problem was to deal separately with the
"unperturbed" rotation curve and the residual velocity field (obtained
supposedly upon subtracting the said curve from the whole picture).
But the problem thus reformulated resisted all attempts to solve it in a
direct way -- in fact, it appears to be ill-conditioned. Indeed, to solve
the problem as stated, one would have to know with high accuracy the
unperturbed velocity at a given radius, which, in turn, is impossible without
the knowledge of the perturbation velocities. The truth is, the problems of
determining the rotation velocity and the perturbations cannot be
separated, because the perturbations caused by a density wave, being
non-random, affect the shape of the restored rotation curve  (see below
Eqs.\ref{eq:purerot}, \ref{eq:AvarphiBrVrot1} and discussion).
The strong interdependency of both parts of the problem is where the
standard methods lose their ground (L97; F97).

\begin{figure}
\epsfig{figure=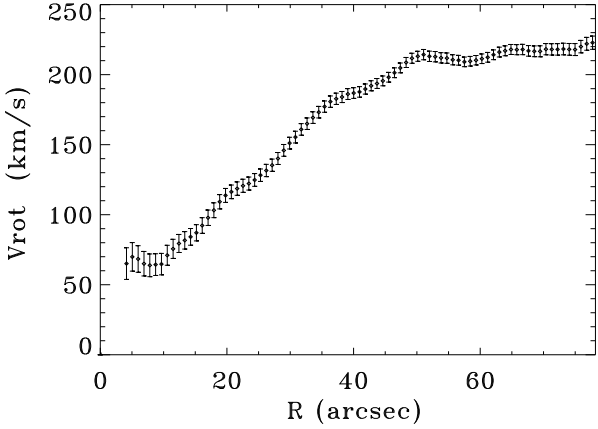,width=8.0cm,
bbllx=85pt, bblly=370pt, bburx=530pt, bbury=690pt, clip=}
\caption[]{The rotation curve of NGC~157 obtained in the frame of the
pure circular motion model for a set of fixed parameters of the
galactic disk. Bars show errors  at the level of 3 $\sigma$.}
\label{f-3}
\end{figure}

As the first step, we shall find the rotation velocity curve of the galaxy
assuming the perturbations to be negligible. It helps to determine the
deviations from pure rotation, to see whether their behaviour
is systematic or random, and in the former case to analyse the connection
of the deviations with the spiral pattern.

In the frame of this model the observed line-of-sight velocity
can be written as (L97):
\[
V^{\rm obs}~=~V^{\rm mod1}~+~ \Psi^{\rm mod1} \, ,
\]
\begin{equation}
\label{eq:purerot}
V^{\rm mod1}(r,\varphi ) ~\equiv~ V^{\rm mod1}_s ~+~ V^{\rm
mod1}_{\rm rot}(r) \cos \varphi \sin i \, ,
\end{equation}
where $\Psi^{\rm mod1}$ is the deviation of the observed velocity
from the one predicted in the model of pure circular motion; $V_s$ is a
systemic line-of-sight velocity of the galaxy;  $V^{\rm mod1}_{\rm rot}$ is the
model circular velocity of the gas in the plane of the galaxy which
depends on the galactocentric radius $r$; $\varphi$ is the galactocentric
azimuthal angle measured with respect to the major axis; $i$ is the
inclination (the angle between the galactic plane and the sky plane).  The
usage of Eq.\ref{eq:purerot} gives the real rotation velocity of the
galaxy if deviations $\Psi^{\rm mod1}$ are random.  Otherwise the value
$V^{\rm mod1}_{\rm rot}$ will differ systematically from the true equilibrium
rotation velocity $V^{eq}_{\rm rot}$ (L97, F97, see also Eqs.\ref{eq:purerot},
\ref{eq:AvarphiBrVrot1} and Fig.~\ref{fig11} below).

For the sake of definition,  hereafter we identify the position $\varphi$
$=$ 0 with the major semiaxis in the receding half of the galaxy and
suppose the inclination to be negative if the angular rotation momentum is
directed toward the observer.  In this work we have assumed that the
center coordinates, the inclination, and the line-of-nodes position angle
are constant parameters which do not depend on the radius.  Note that
variations of the disk orientation parameters along the radius (often
obtained as a result of the "splitting into rings" method of calculation
of the said parameters, when each ring is considered independently), may
be an artefact arising from neglecting the systematic deviations of the
gas velocities from the circular ones (L97).

To determine the orientation parameters of the rotation plane of NGC~157 and
its rotation curve, we have used more than 11~000 individual velocity
measurements\footnote{It might be useful to emphasise here the difference
between "spatially resolved" and "statistically independent" observational
points. In our case, the latter correspond to "individual velocity
measurements", numbering up to 11~000. Evidently, the
number of the "spatially resolved" observational points depends on seeing,
while the number of the "statistically independent" observational points does
not.}. The model parameters for the galaxy were calculated by the method
of least squares, minimizing the r.m.s.  deviation of the observed
velocities from the model ones derived from equation (\ref{eq:purerot}).
To estimate the rotation velocity at a radius $r$, we used the line-of-sight
velocity values in the range $r- d/2$ to $r+ d/2$.  The value of the ring
width $d$ was chosen on the basis of the following arguments.
Increasing $d$ decreases the random errors, but at the same time it
leads to the growth of the systematic error caused by neglecting
gradients in the rotation velocity. The latter error could be
estimated as $\varepsilon_{sys}$ $\approx$ $V_{\rm rot}''\,d^2/8$.  Thus
demanding the systematic error to be less than 2 km s$^{-1}$ and
evaluating $V''_{\rm rot}$ $\approx$ (200~km s$^{-1}$)/(20$'')^2$ we come
to the restriction on $d$ from above:  $d$ $<$ 6$''$. On the other hand,
it is meaningless to choose $d$ smaller than  the seeing value (2.5$''$).
For these reasons, in the analysis presented below we use $d$ $=$ 5 pixels
or 4.6 arcsec. The rotation velocity was calculated at the step of one
pixel (0.92 arcsec).  After running through the whole range of the radial
distances, we fit the rotation curve by a cubic spline.

\begin{figure}
\epsfig{figure=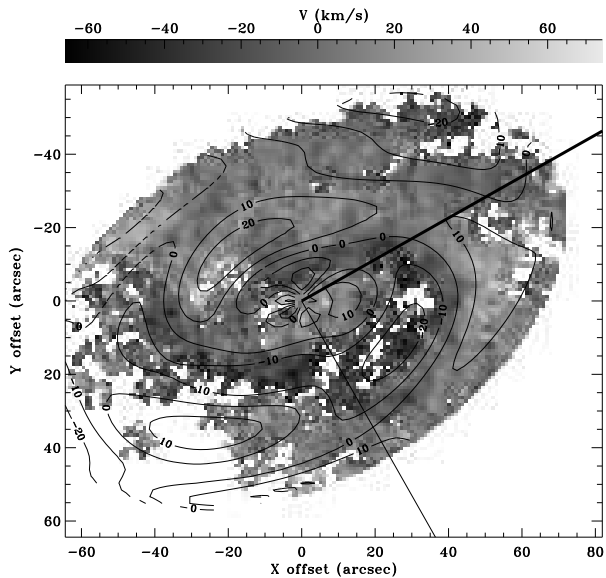,width=8.0cm,
bbllx=75pt, bblly=138pt, bburx=534pt, bbury=577pt, clip=}
\caption[]{The field of the residual line-of-sight velocities in NGC~157
with isovelocity contours overlaid. One can see that the deviations from
the pure circular motion demonstrate a systematic character.} \label{f-4}
\end{figure}

As a result, we have obtained the following best fit parameters for
NGC~157:  the inclination $i = -51.5^\circ\pm 5^\circ$, the position angle
of the major axis (line of nodes) PA$_0 = 223.5^\circ\pm 2^\circ$ (we
identify $PA_0$ with the orientation $\varphi=$ 0), and the systemic
heliocentric velocity of the galaxy $V_{sys} = 1667 \pm  5$~km s$^{-1}$.
For comparison, in the RC3 (de Vaucouleurs et al. 1991) one can
find the following estimates of the systemic velocity for NGC~157: $1730
\pm 27$ km s$^{-1}$ from  optical data and $1668 \pm 6$ km s$^{-1}$
from radio (H{\sc i} line) data.  Among the above listed numerical
parameters the value of $i$ was estimated less reliably.  From the
isophote ellipticity the inclination is known to be $i = 45^\circ -
48^\circ$ (Grosb{\o}l 1985; Bottinelli et al. 1984).  In the paper of
Ryder et al. (1998), where the same kinematical data were used for the
inner part of the galaxy and a more traditional method was
applied to find the disk parameters, the inclination was found to be $i =
45^\circ \pm 5^\circ$.  Thus within the limits of observational errors our
determination of the inclination $i$ agrees with the results of other
authors.

The obtained model rotation curve $V^{\rm mod1}_{\rm rot}$ is presented in
Fig.~\ref{f-3}.  The errors of the rotation velocity (3 $\sigma$) are
shown by bars and are of order of 3~km s$^{-1}$ (this value represents
random errors only, systematic errors caused by the neglecting of motions
in the density wave could be much higher).  One can see that over a wide
radial range ($1 - 5$ kpc) $V^{\rm mod1}_{\rm rot}$ rises quite linearly with
the radius $r$; the fact was pointed out earlier by Barbridge et al. (1961)
and Zasov \& Kyazumov (1981).  This peculiarity distinguishes NGC~157
from the majority of the other spiral galaxies, for which such regions
of monotonous rise of the rotation curves rarely extend beyond $2 - 3$ kpc.

\section{The model taking into account motions in the spiral arms}

In Fig.~\ref{f-4} we present a map of the gas residual velocities (the
difference between the observed velocities and the model ones obtained
from (\ref{eq:purerot})). As one can see, the observed velocities deviate
systematically from the pure circular motion. The regions of maximum
deviations have a spiral shape, which proves their connection
with the spiral density wave in this galaxy. It means that the assumptions
underlying the model of pure circular motions are not
valid and a more detailed model should be build.

To construct a model describing the velocity field of
NGC~157 more appropriately, it is necessary to take into account
motions in the density wave. In the general case, the equation,
connecting the components of the full velocity vector in the
galactocentric frame,  $V_r (r,\varphi )$, $V_\varphi(r,\varphi )$, and
$V_z (r,\varphi)$, with the observed line-of-sight velocity may be
written as follows (L97; F97):
\[
V^{\rm obs}(r, \, \varphi) ~=~ V_s ~+~ V_\varphi(r, \, \varphi ) \, \cos
\varphi \, \sin i ~+~
\]
\begin{equation}
\label{eq:vobsa}
~+~ V_r (r, \, \varphi) \, \sin \varphi \, \sin i ~+~ V_z (r, \, \varphi)
\,\cos i ~+~ \Psi^{\rm obs} \, ,
\end{equation}
where $\Psi^{\rm obs}$ are errors in  the observations. Values
$V^{\rm obs}$, $V_s$, $V_r$, $V_\varphi$ and $V_z$ describe the velocity
field at the moment of the observations. In other words, they are
functions of the galactocentric radius $r$ and the azimuth $\varphi$ but
do not depend on the time. In the case of pure circular motion,
$V_\varphi(r,\,\varphi)$ $=$ $V^{\rm mod1}_{\rm rot}(r)$,
$V_r(r,\,\varphi)$ $=$ 0, $V_z(r,\,\varphi)$ $=$ 0, so this formula
evidently reduces to (\ref{eq:purerot}).

The relation (\ref{eq:vobsa}) is the only one between the line-of-sight
velocity field and the components of the vector velocity field of the
gaseous disk, which can be derived without loss of generality. As a
result, the direct restoration of the velocity field of the gaseous disk
from the line-of-sight velocity field is impossible and some indirect
methods should be applied (F97).

\begin{figure}
\epsfig{figure=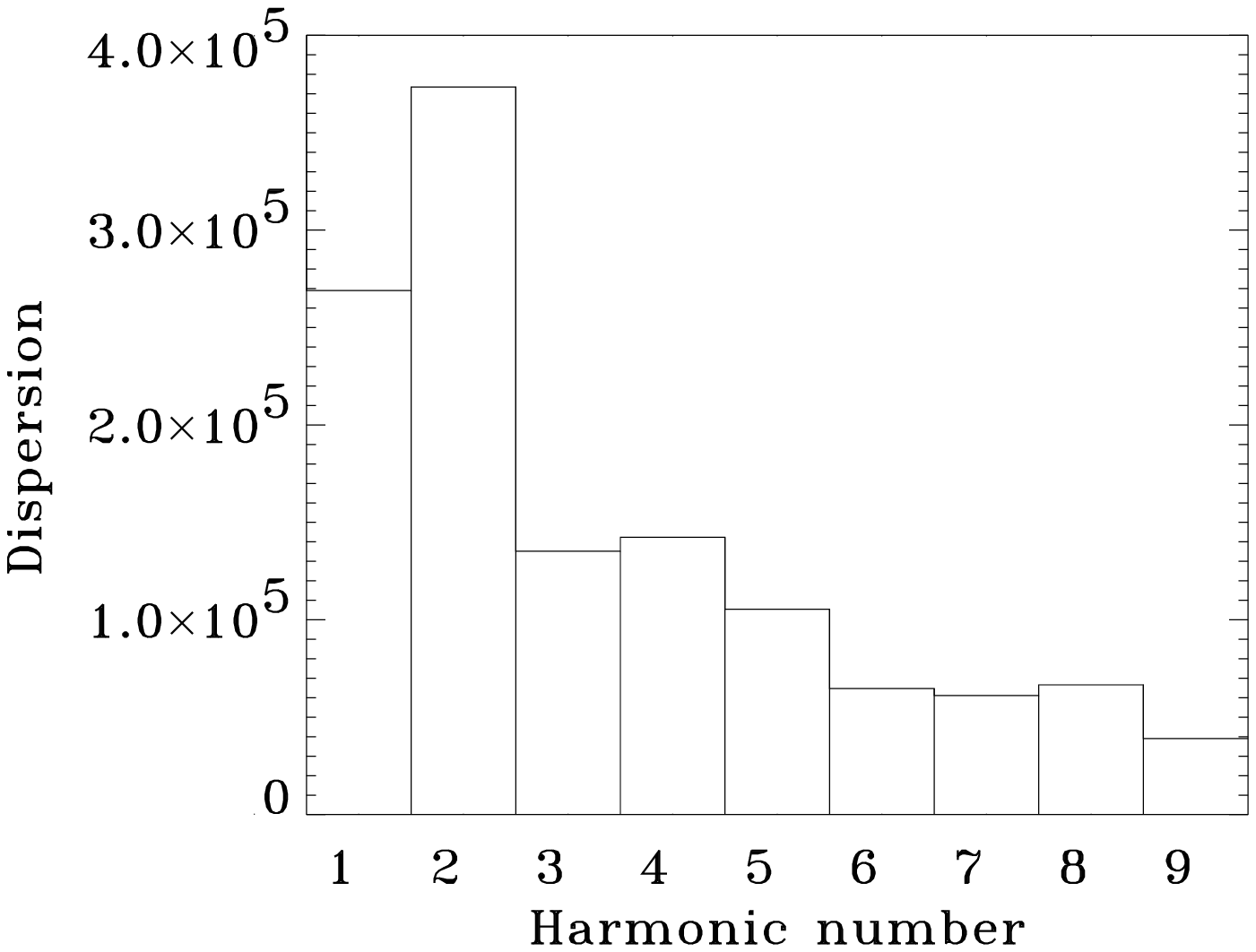,width=8.0cm,
bbllx=80pt, bblly=70pt, bburx=485pt, bbury=380pt, clip=}
\vspace{-.4cm}

~~a)

\epsfig{figure=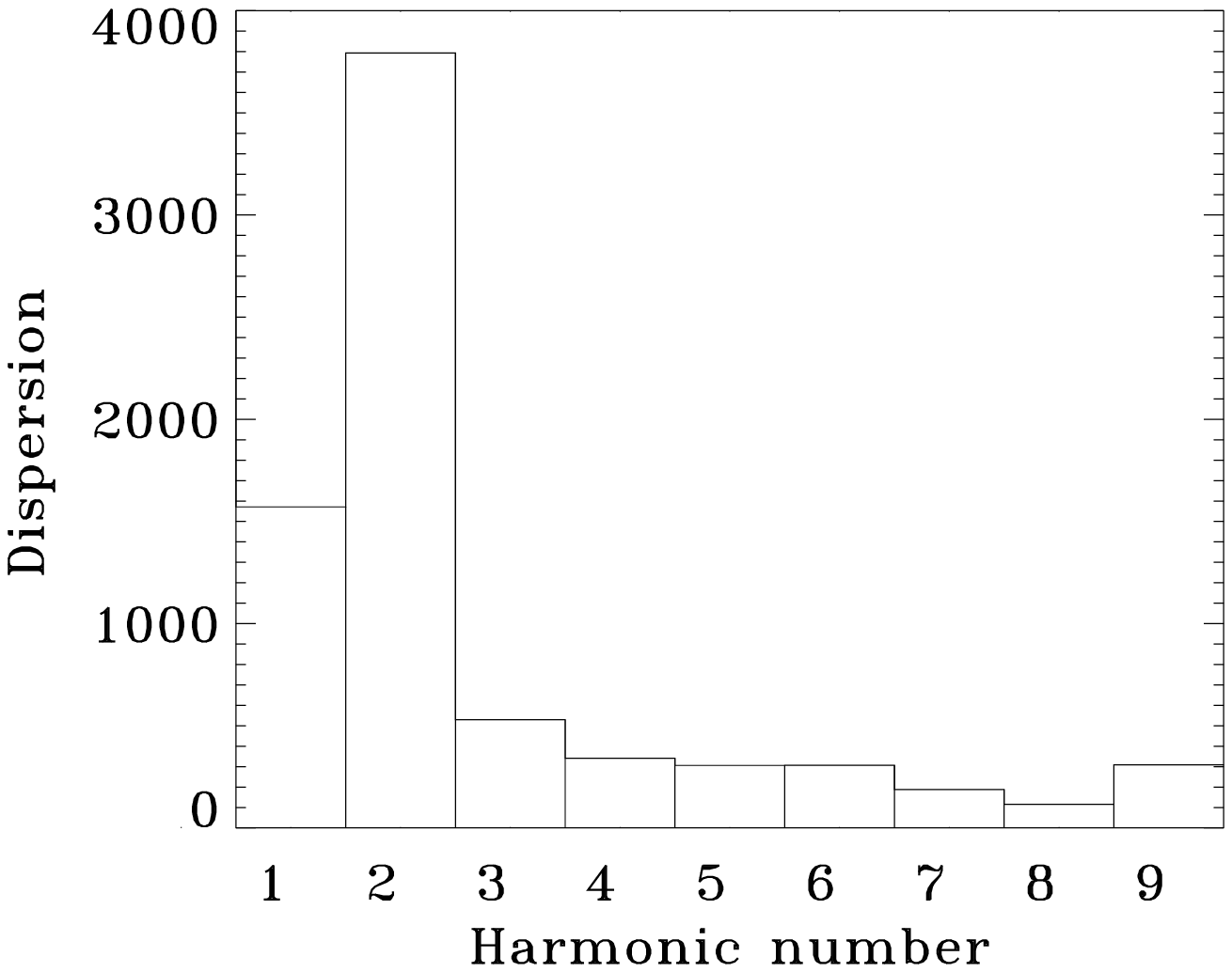,width=8.0cm,
bbllx=80pt, bblly=70pt, bburx=485pt, bbury=380pt, clip=}
\vspace{-.4cm}

~~b)

\caption{Averaged over the radial range 5$''$ - $75''$ the
contribution (dispersion given in arbitrary units) of the individual
Fourier harmonics into the deviation of the brightness distribution in
different bands from the azimuthal symmetry. (a) H$\alpha$ brightness
map, (b) $K$-band brightness map (kindly provided by S.D.~Ryder).}
\label{f-denaa}
\end{figure}

Particularly, one may pose the question in the following way. Let us take
into account that we study a galaxy with clear two--armed spiral structure
(for the general case of an $m$--armed spiral see L97; F97). As the
arms are easy to trace, one may assume that the perturbed surface density
of the gaseous disk $\tilde \sigma$ (at the moment of observation) can be
approximated by the expression
\begin{equation}
\tilde \sigma ~=~C_\sigma(r)~ \cos [ 2 \varphi - F_\sigma(r)] \, ,
\label{vst2}
\end{equation}
where $C_\sigma$ and $F_\sigma$ are the corresponding amplitude and phase.

The validity of using (\ref{vst2}) may be checked by Fourier analysis of
the optical image of the galaxy: the approach is valid if the amplitude of
the second Fourier harmonic  predominates over the others and its phase curve
closely correlates with the observed spiral.  As one can see from
Fig.~\ref{f-denaa}, the second harmonic dominates.  The high level of the
first harmonic in the H$\alpha$ brightness map is caused by the
nonsymmetrical distribution of star forming regions in the observed spiral
arms. In our opinion, it does not reflect the real contribution of the
first Fourier harmonic into the distribution of the perturbed surface
density in the gaseous disk of the galaxy. The relatively low level of the
first harmonics in the $K$-band image proves this assumption.
Fig.~\ref{f-2} demonstrates that for NGC~157 the phase curve of the second
Fourier harmonic of the H$\alpha$ brightness map is in good accordance
with the observed spiral pattern as well as with the second harmonic of
$K$-band image of the galaxy.

In a frame of the approach described, the perturbed velocity components
may be written in a similar manner:
\begin{equation} \label{eq:sprvz1}
\tilde{V_r}(r,\,\varphi ) = C_r(r) \, \cos [2\varphi- F_r (r) ] \, ,
\end{equation}
\begin{equation}
\label{eq:sprvz2}
\tilde{V_\varphi}(r,\,\varphi ) = C_\varphi(r) \, \cos  [ 2\varphi-
F_\varphi (r) ] \, ,
\end{equation}
\begin{equation}
\label{eq:sprvz3}
\tilde{V_z}(r,\,\varphi ) = C_z(r) \, \cos  [2\varphi - F_z (r)] \, .
\end{equation}

The validity of this approximation also can be checked directly from the
observations.  Indeed, taking into account that $V_r(r,\,\varphi )$ $=$
$\tilde{V_r}(r,\,\varphi )$, $V_\varphi (r,\,\varphi )$ $=$
$V^{\rm mod2}_{\rm rot}(r)$ $+$ $\tilde {V_\varphi} (r,\,\varphi)$,
$V_z(r,\,\varphi )$ $=$ $\tilde{V_z}(r,\,\varphi )$, and substituting
(\ref{eq:sprvz1}) -- (\ref{eq:sprvz3}) in (\ref{eq:vobsa}), we obtain the
model representation of the line-of-sight velocity (L97, F97):
\[
V^{\rm mod2}(r,\,\varphi ) ~=~ V^{\rm mod2}_s ~+~ \sin i \, \left[\,
a^{\rm mod2}_{1}(r) \, \cos \varphi ~+~ \right.
\]
\[
+~b^{\rm mod2}_{1}(r) \, \sin \varphi + a^{\rm mod2}_2(r) \, \cos 2\varphi
+ b^{\rm mod2}_2(r) \, \sin 2 \varphi~ +
\]
\begin{equation}
\label{eq:vobsout20}
\left.~~~~~~~+~ a^{\rm mod2}_{3} (r) \, \cos 3\varphi ~+~  b^{\rm mod2}_{3} (r) \,
\sin 3 \varphi \, \right] \, ,
\end{equation}
where Fourier coefficients related to the phases and amplitudes of the
velocity components are:
\begin{equation}
\label{eq:aAbB1}
a^{\rm mod2}_{1} ~=~ V^{\rm mod2}_{\rm rot}(r) ~+~ \frac {C_r \sin F_r + C_\varphi \cos
F_\varphi  } 2 \, ,
\end{equation}
\begin{equation}
\label{eq:aAbB2}
b^{\rm mod2}_{1} ~=~ -~ \frac{C_r \cos F_r - C_\varphi \sin F_\varphi }2 \, ,
\end{equation}
\begin{equation}
\label{eq:aAbB3}
a^{\rm mod2}_2 ~=~ C_z \cos F_z \,{\rm cot}\, i \, ,
\end{equation}
\begin{equation}
\label{eq:aAbB4}
b^{\rm mod2}_2 ~=~ C_z \sin F_z \,{\rm cot}\, i \, ,
\end{equation}
\begin{equation}
\label{eq:aAbB5}
a^{\rm mod2}_{3} ~=~ -~ \frac{C_r \sin F_r - C_\varphi \cos F_\varphi} 2 \,  ,
\end{equation}
\begin{equation}
\label{eq:aAbB6}
b^{\rm mod2}_{3}~ =~ \frac{C_r \cos F_r + C_\varphi \sin F_\varphi }2 \,  .
\end{equation}

From the above relationships (\ref{eq:aAbB1}) - (\ref{eq:aAbB6}) we can
see that the contributions of the different velocity components in the
azimuthal Fourier harmonics of the observed line-of-sight velocity are
distributed as follows.
\begin{enumerate}
\item The systemic velocity of the galaxy contributes in the zeroth
harmonic of the observed line-of-sight velocity.
\item The circular rotation contributes in the coefficient of cosine of
the  first component of the observed line-of-sight velocity.
\item The perturbed motion in the galactic plane contributes in the first
and third harmonics of the observed line-of-sight velocity.
\item The vertical motion (along the rotation axis) contributes in the
second  harmonic of the observed line-of-sight
velocity\footnote{Hereafter, by vertical motions we
mean, first of all, the vertical motions in the density wave. The vertical
velocity component in the density wave is antisymmetrical with respect to
the central plane of the disk and thus leaves the disk flat on average.
These motions are observable due to the non-transparency of the disk
in the H$\alpha$ line. As a result, the contribution of the nearest part
of the disk prevails. The connection of the observed vertical motions with
the density wave can be checked analysing the correlation between phases
of the second harmonic of the brightness map with the second harmonic of
the line-of-sight velocity field (Sec.~5.3) or comparing amplitudes of the
second harmonics of the line-of-sight velocity fields obtained in lines
with different optical width (the method proposed in Fridman et al.
1998).}.
\end{enumerate}

\begin{figure}
\epsfig{figure=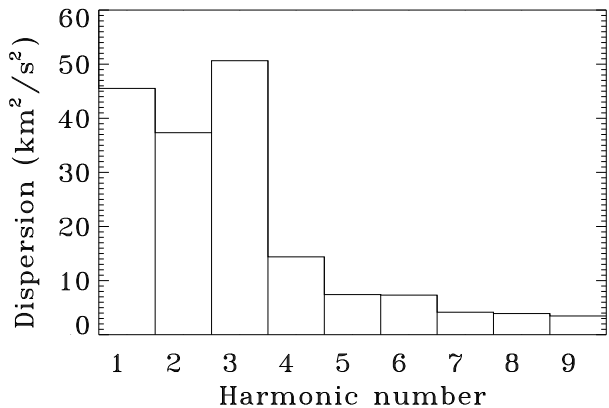,angle=0,width=8.0cm,
bbllx=125pt, bblly=355pt, bburx=485pt, bbury=600pt, clip=}
\caption[]{Averaged over the radial range 5$''$ - $75''$, the contribution
of the individual Fourier harmonics into dispersion  in the model
of pure rotation. The prevalence of the first sine, second and third
harmonics is clearly seen.}
\label{f-5}
\end{figure}

Thus, if a galaxy has a pure two--armed structure, the observed
line-of-sight velocity field should contain the zeroth, the first, the
second and the third harmonics only.  Actually, if the structure is not
pure two--armed as we have in the case of NGC~157 (see Sandage \& Bedke
1994), and yet the two--armed mode dominates, one can expect a
dominance of these harmonics in the observed line-of-sight
velocity field (F97).

Fig.~\ref{f-5} shows, for NGC~157, the contributions of the
different Fourier harmonics to dispersion in the model of pure rotation
averaged over the radius range 5$''$ - 75$''$. It is clearly seen that the
deviations are mainly determined by the first (sine), second, and third
harmonics. The individual contributions of any higher harmonic are
relatively small.

Hence, the observations of NGC~157 within the radial range
mentioned above justify the possibility to restrict the input of motions
induced by the density wave to the main ($2\varphi$) component only. It
allows the use of relations (\ref{eq:aAbB1}) --
(\ref{eq:aAbB6}) connecting characteristics of the perturbed velocities
and Fourier coefficients of line-of-sight velocity field to restore the
three-dimensional velocity field of the galaxy.

The results obtained in this way will be presented in
Sec.~\ref{sec-6}.  But before proceeding, we should check whether, indeed,
there exists a close connection between the residual velocities and
the spiral arms observed in NGC~157. So, let us turn to the Fourier
analysis of the line-of-sight velocity field to prove the wave nature of
the spiral structure and the observed velocity deviations from pure
rotation.

\section{Fourier analysis of the velocity field}

\subsection{Fourier analysis technique}

Suppose that the inclination $i$, the position angle of the
kinematical major axis (the line of nodes) PA, and the coordinates of the
galactic center in the sky plane $x_0$ and $y_0$ are known (say, having
been determined in  some independent way).

If the observed line-of-sight velocity can be approximated by a
two-dimensional analytical function of galactocentric coordinates $r$
and $\varphi$, then its expansion into a harmonic series may be
presented as:
\begin{equation}
\label{eq:obsfur}
V^{\rm obs} = V_{s} + \sin i \sum_{n=1}^{n_{max}} \left( a^{\rm obs}_n (r) \cos
n\varphi + b^{\rm obs}_n (r) \sin n\varphi \right)
\end{equation}
where
$n_{max}$ is the number of the highest harmonic, to be chosen for each
galaxy individually.

A well--known technique has been applied to find Fourier coefficients at
discrete points $r= R_l$ by minimising the difference
\[
\chi^2( R_l)=\sum_{j} \left(V_j (r_j,\varphi_j)- V_{s} -\right.
\]
\begin{equation}
\label{eq:leastsqr1}
\left.~-~\sum_{n=1}^{n_{max}} \left( a^{\rm obs}_n ( R_l) \cos n\varphi_j
+b^{\rm obs}_n ( R_l) \sin n\varphi_j \right) \sin i \right)^2 \,.
\end{equation}
To construct r. m. s. deviations $\chi^2 ( R_l)$ we use the observed
values of $V_{j}$  with the coordinates $r_j$ falling in the
galactocentric radius range from $R_l-d/2$ to $ R_l+d/2$ with $d$ $=$ 4.6
arcsec (see discussion on the choice of $d$ in Sec.3).  Finally, to find
$a^{\rm obs}_n(r)$ and $b^{\rm obs}_n(r)$ in intermediate points $r$ we
use the cubic spline drawn through the calculated points $a^{\rm obs}_n(
R_l)$ and $b^{\rm obs}_n( R_l)$.

The recent investigation by Burlak et al. (2001) has shown that the method
described here is very stable, and that its results are practically
insensitive to the presence of 'holes' in the velocity fields if the
filling factor of points with velocity measurements for a given
galactocentric ring is greater than 30\%.

\subsection{The large-scale line-of-sight velocity field}

The square root of the average dispersion in the model of pure rotation is
about 19 km s$^{-1}$, whereas the random error of a velocity measurement
(in one pixel) is about 14 km s$^{-1}$. As the error ($\sigma$) of the
harmonic amplitude determination is about 1.5 km s$^{-1}$, the above
difference is statistically significant on the level of 99\%. As mentioned
above, deviation from the pure rotation is mainly determined by the first
(sine), second and third harmonics.

\begin{figure}
\epsfig{figure=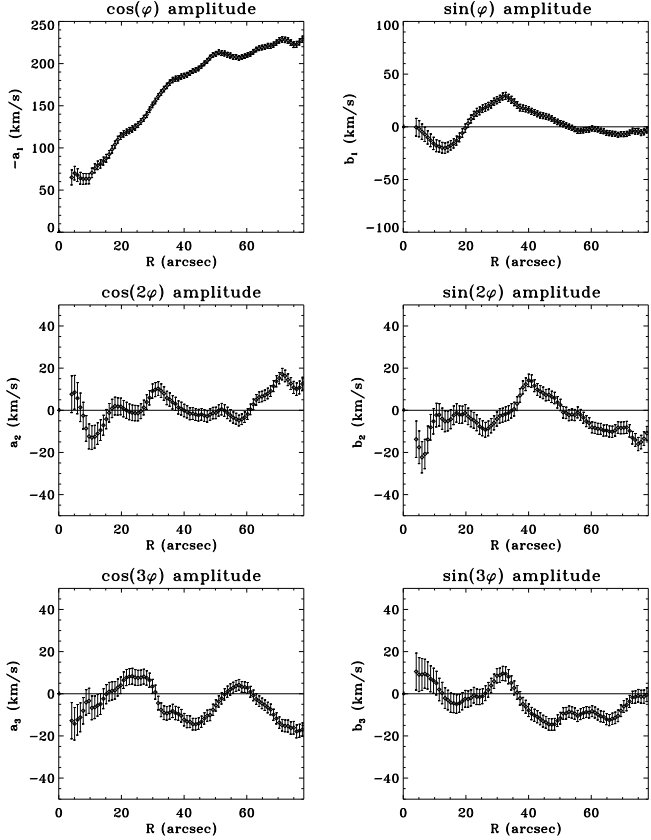,width=8.6cm,
bbllx=99pt, bblly=90pt, bburx=553pt, bbury=680pt, clip=}
\caption[]{Radial behaviour of the main harmonics of the line-of-sight
velocity field for NGC~157. Bars show the observational errors
(3$\sigma$).}
\label{f-6}
\end{figure}

The behaviour of the main harmonics with galactocentric radius is shown
in Fig.~\ref{f-6}.  Bars correspond to 3$\,\sigma$ level. This figure
demonstrates that the amplitudes of the main harmonics exceed the errors
substantially, and hence, their values can be determined from the
observations reliably over most of the galactic disk. The first
plot shows the absolute value of the first harmonic cosine
coefficient\footnote{The proper value of the coefficient $a^{\rm obs}_1$ is
negative, which corresponds to counterclockwise galaxy rotation from the
observer's point of view.}. Due to the good filling of the galactic
disk by the velocity measurements, the value of $a^{\rm obs}_1$ is close
to the value calculated with no account for the contribution of the other
Fourier harmonics, that is by the reverse Fourier
transformation\footnote{The latter value is equal to rotation velocity
$V_{\rm rot}^{\rm mod1}$ calculated in the model of pure circular
motions.}. A typical discrepancy does not exceed 1 km s$^{-1}$, which is
less than the errors of the $a^{\rm obs}_1$ determination. The maximum
difference (of the order of 5 km s$^{-1}$) takes place near the galactic
center where there are only a few measurements suitable for the Fourier
analysis, and at the edge of the galaxy where "holes" exist in the
observed velocity field. It should be noted that the mentioned closeness
is a direct consequence of the independence of different Fourier harmonics
and bears no relation to the validity of the model of pure circular
motions. According to Eq.~(\ref{eq:aAbB1}), the difference between $a_1$
and the equlibrium rotation velocity is systematicaly nonzero and about
the amplitude of motions in the density wave (i.e about 10-20 km
s$^{-1}$, see Fig.~\ref{fig11} below).

Now we restrict the analysed field of the line-of-sight velocity of the
galaxy only to the first, second, and third harmonics:
\[
V^{red}/\sin i ~=~ V_{sys}/\sin i ~+~
\]
\[
~+~a_1^{\rm obs} (r) \cos \varphi ~+~
b_1^{\rm obs} (R) \sin \varphi ~+~
\]
\[
a_2^{\rm obs} (r) \cos 2\varphi ~+~ b_2^{\rm obs}(r) \sin 2\varphi ~+~
\]
\begin{equation}
\label{eq:furlast}
~+~a_3^{\rm obs}
(r) \cos 3\varphi ~+~ b_3^{\rm obs} (r) \sin 3\varphi \,.
\end{equation}
The averaged velocity dispersion in the model is about 15~km s$^{-1}$. The
restriction to the first three harmonics is reasonable from the
goodness-of-fit point of view (see Bevington 1975) because it enables
the elimination of systematic deviations of the observed velocity
field from those expected in the case of pure rotation of the gas in
the plane of the disk.

\begin{figure}
\epsfig{figure=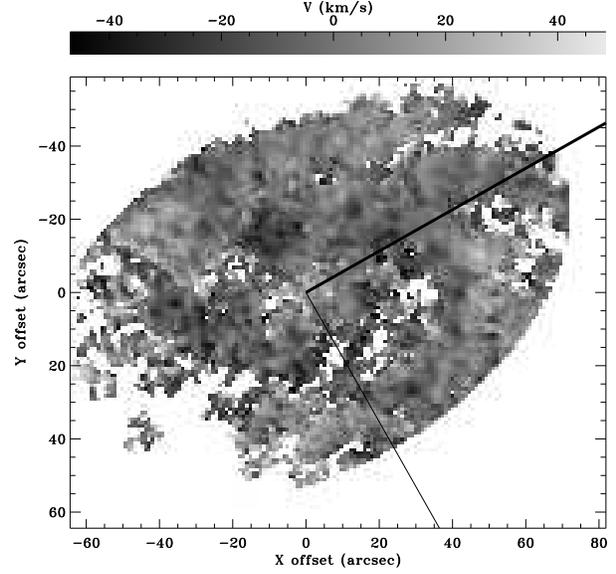,width=8.0cm,
bbllx=75pt, bblly=138pt, bburx=535pt, bbury=574pt, clip=}
\caption[]{Deviations of the observed line-of-sight velocity field
from that restricted to the first three Fourier harmonics.  The chaotic
small-scale structure of the field is clearly seen. It is caused by random
errors and by small-scale motions in the star-formation regions.}
\label{f-9}
\end{figure}

Fig.~\ref{f-9} presents the residual velocity field for this model -- the
observed velocity field  minus the model one, restricted to the first
three Fourier harmonics according to (\ref{eq:furlast}). As one can see,
the residual velocities demonstrate a small-scale chaotic structure.
The areas of maximum deviations correlate with the intense star-formation
regions seen on the H$\alpha$ image of the galaxy (Fig.~\ref{f-2}).

Therefore we can conclude that the first three harmonics describe  the
large-scale structure of the velocity field rather well, while the
higher-order harmonics over the major part of the galaxy have smaller
amplitudes and relate mostly to small-scale distortions in the brightest
star formation regions and to random errors.  Hence for further
analysis of the large-scale structure of the velocity field we can
restrict ourselves to the first three harmonics.

In the frame of the three--harmonic model we have reevaluated the
parameters of the orientation of the galactic rotation plane ($i$ and
PA$_0$). Our calculations show that the parameters which minimize the r.
m. s.  deviations of the line-of-sight velocities from (\ref{eq:furlast})
coincide with those found in the frame of the pure circular model within
the measurement errors. This  is easy to explain, remembering that
the harmonic coefficients oscillate along the radius (see Fig.~\ref{f-6}),
so, being averaged, their input is close to zero.

Therefore, for a given galaxy and for a given rich statistical sample
of measurements of the line-of-sight velocity, the model of pure
circular motion allows to find parameters of the orientation of the
disk with a good accuracy, even without taking into account the motions
in the spiral arms.

\begin{figure}
\epsfig{figure=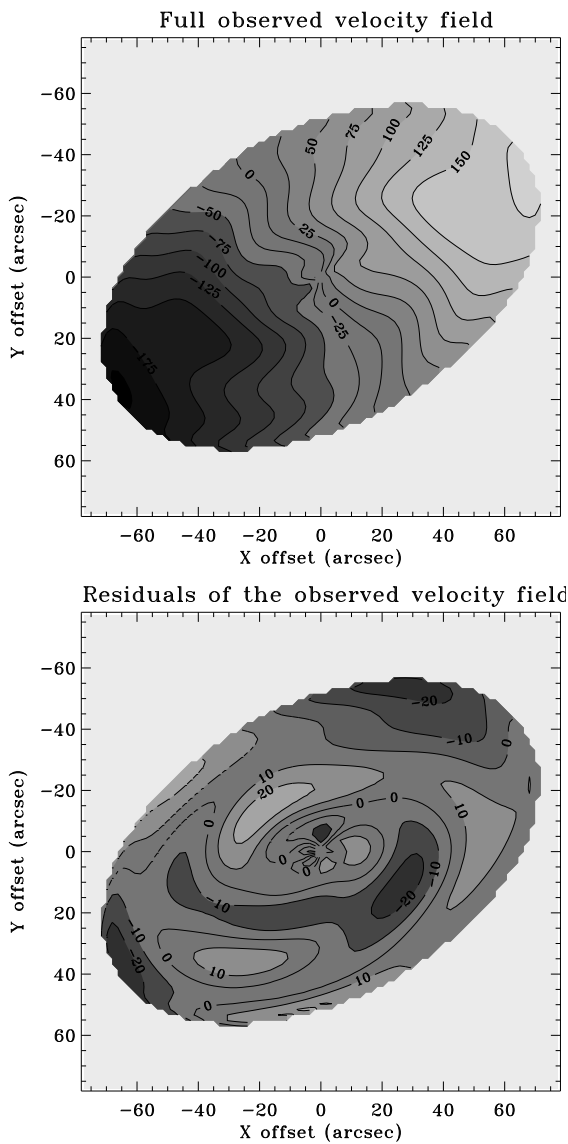,width=7.5cm,
bbllx=95pt, bblly=95pt, bburx=430pt, bbury=770pt, clip=}
\caption[]{Large-scale component of the full (top) and residual
(bottom) line-of-sight velocity fields. Only contributions of
the first, second and third Fourier harmonics are included.}
\label{f-10}
\end{figure}

Fig.~\ref{f-10} (top) shows a model velocity field of NGC~157 in the
sky plane, calculated by using the formula (\ref{eq:furlast}).
Fig.~\ref{f-10} (bottom) shows a large-scale residual velocity field which
represents the model velocity field upon subtraction of the first cosine
harmonic. The picture of this residual velocity field is rather
complex.

\begin{figure}
\epsfig{figure=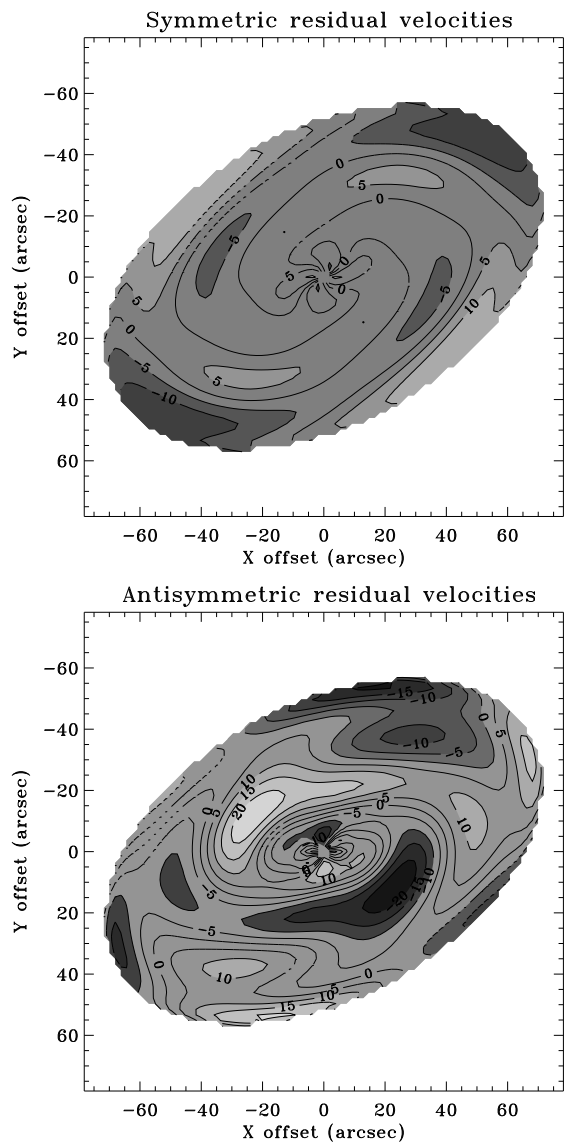,width=7.5cm,
bbllx=95pt, bblly=95pt, bburx=430pt, bbury=770pt, clip=}
\caption[]{The symmetric (top) and antisymmetric (bottom) parts
of the large-scale residual line-of-sight velocity field.}
\label{f-11}
\end{figure}

To clarify the structure of the velocity field, in Fig.~\ref{f-11} we
present a symmetric (top) and antisymmetric (bottom) parts
of the residual velocity field separately. The symmetric part contains the
contribution of the second harmonic, and the antisymmetric part --
contributions of the third harmonic and of the first sine harmonic.

Note that although Fig.~\ref{f-11} (bottom) resembles the residual
velocity maps of Canzian (1993), use of Canzian's method for
the determination of the corotation radius became possible only
after separating the residual velocity field into symmetric and
antisymmetric components which  has not been done by Canzian.  As
for the original field of residual velocities (Fig.~\ref{f-10}, bottom),
its shape is influenced by the second Fourier harmonic and does not
contain any switching from one-armed to three-armed spiral, that is, does
not allow use of Canzian's method in principle. It is also worth
mentioning that Canzian (1993) did not take the second harmonic into
account at all.

\subsection{Proof of the wave nature of the grand design arms}

As mentioned above, the dominance of the first, second and third
harmonics in the expansion of the line-of-sight velocity field of NGC~157
into Fourier series finds a natural explanation under the assumption of
the wave nature of the grand design structure (see also F97).

\begin{figure}
\epsfig{figure=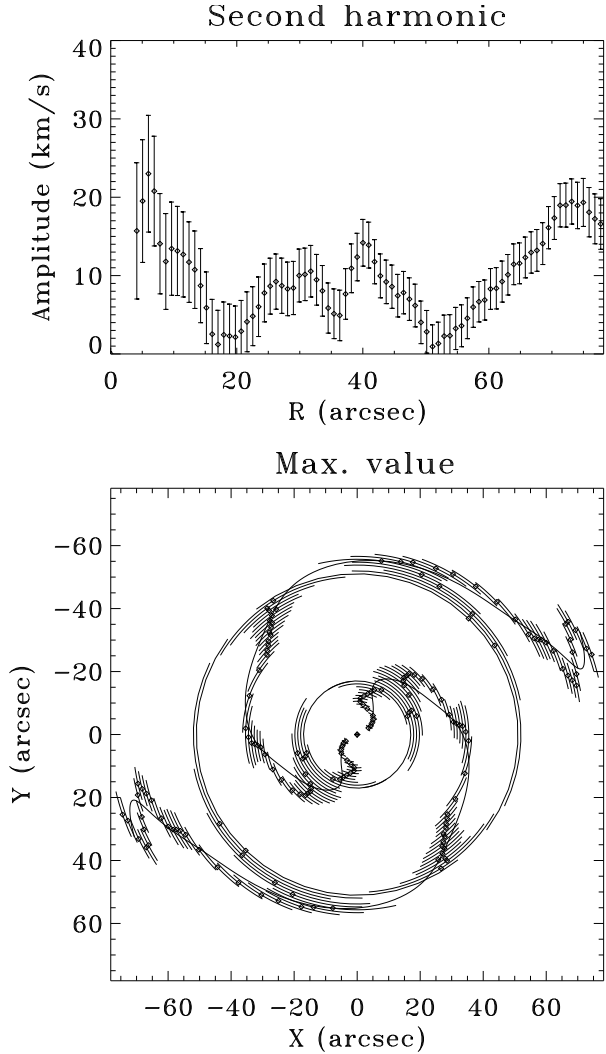,width=8.cm,
bbllx=88pt, bblly=120pt, bburx=470pt, bbury=790pt, clip=}
\caption[]{Radial behaviour of the amplitude (top) and the phase (bottom)
of the second Fourier harmonic of the
line-of-sight velocity. Bars show the observational errors (3$\sigma$).}
\label{f-7_2}
\end{figure}

Let us obtain direct evidence for the latter on the base of the
connection between the main Fourier harmonics of the line-of-sight
velocity field and the spiral structure observed in NGC~157. To do this we
consider in detail the behaviour of the second and the third Fourier
harmonics. For further consideration it is convenient to define the
amplitude $C_n$ and the phase $F_n$ for any harmonic by the following
equations:
\begin{equation}
\label{vst14}
C_n\,\cos(F_n) ~=~ a_n^{\rm obs} \,,
\end{equation}
\begin{equation}
C_n \, \sin(F_n)~=~b_n^{\rm obs} \,.
\label{vst15}
\end{equation}

\begin{figure}
\epsfig{figure=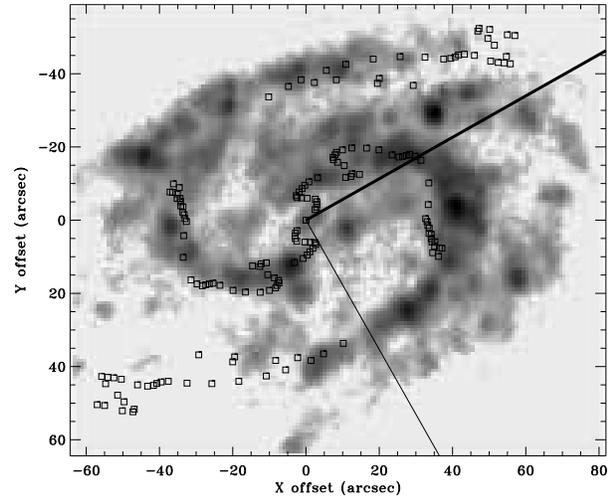,width=8.0cm,
bbllx=75pt, bblly=190pt, bburx=535pt, bbury=575pt, clip=}
\caption[]{Superposition of the second harmonic of the line-of-sight
velocity field on the H$\alpha$ image of NGC~157 (gray scale). Triangles
show the azimuth positions of the maxima of the harmonic at each radius.
The good correspondence of the phase curve with the observed
position of the spiral arms gives strong evidence for the non-circular
velocities, associated with the spiral structure.}
\label{f-8}
\end{figure}

Fig.~\ref{f-7_2}  presents radial variations of the amplitude
and phase for the  second harmonic of the line-of-sight
velocity, following (\ref{vst14}) and (\ref{vst15}) for $n$ $=$ 2.
The bottom plot shows the locus of the second harmonic
maximum (i.e. where $\cos[2\varphi-F_2(R)]$ $=$ 1) on the galactic plane,
which demonstrates the behaviour of the second harmonic phase.
Arcs on the plot show errors of the phase determination (3 $\sigma$). As
one can see, there are only a few radial positions (near $r$ $=$
$20\arcsec$ and 50\arcsec) where the amplitude of the second
harmonic is comparable with the error of its measurement ($C_2$
$\le$ $3 \cdot \sigma_{c_2}$), while over most of the galaxy it
is determined quite safely ($C_2$ $>$ $3 \cdot \sigma_{c_2}$). The lines
of the second harmonic maxima look like a two-armed trailing spiral, and
they are well correlated with the galactic spiral arms observed in the
emission line of H$\alpha$  inside 50\arcsec (Fig.~\ref{f-8}). This
gives strong evidence that the second harmonic of the line-of-sight
velocity is related to the vertical motions in the observed spiral density
wave in the region $r$ $<$ $50\arcsec$. In the outer region, the second
harmonic of the line-of-sight velocity can be caused by the plane motions
in $m$ $=$ 3 mode, which becomes relatively stronger in this region.
A discussion of this subject is beyond the scope of the current paper.

\begin{figure}
\epsfig{figure=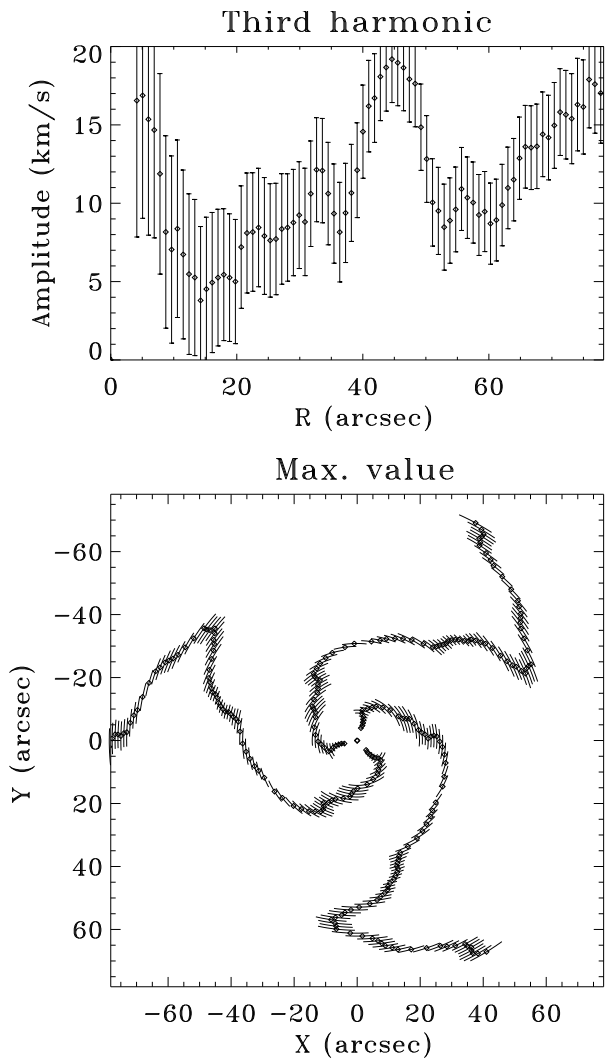,width=8.cm,
bbllx=88pt, bblly=120pt, bburx=470pt, bbury=790pt, clip=}
\caption[]{Radial behaviour of the amplitude (top) and the phase (bottom)
of the third Fourier harmonic of the line-of-sight velocity. Bars show the
observational errors (3$\sigma$).}
\label{f-7_3}
\end{figure}

Fig.~\ref{f-7_3}  presents the behaviour of the amplitude and phase
for the third Fourier harmonic of the line-of-sight velocity field, which
are defined by equations (\ref{vst14}) and (\ref{vst15}) for $n$ $=$ 3,
as well as their errors.  The bottom plot shows the locus which is
determined by the equation $\cos[3\varphi-F_3(R)]$ $=$ 1 on the galactic
plane. The amplitude of the third harmonic over most of the galaxy is
determined rather reliably ($C_3 > 3 \cdot \sigma_{c_3}$).  The lines of
maxima of the third harmonic also look like trailing spirals over most
of the galaxy, excluding a small area at the periphery ($r$ $>$
$60\arcsec$), where they have the  form of the leading spirals.

Though the spiral-like behaviour of the third harmonic is evident,
a  direct comparison of its pattern with the observed spiral arms of
the galaxy is impossible because of the unequal number of arms.
Nevertheless, it can be shown that if the observed spiral structure is a
tightly wound density wave, there must exist a connection between the
phase of the third Fourier harmonic of the line-of-sight velocity field
and the phase of the perturbed surface density (L97; see also the
Appendix below):  \begin{equation}
\label{vst13a}
F_3~\approx~F_\sigma ~+~ \pi/2 \, .
\end{equation}
Within the corotation radius, this relation will be valid if $C_\varphi$
$>$ $C_r$. Under the opposite condition $C_\varphi$ $<$ $C_r$, $F_3$
$\approx$ $F_\sigma-\pi/2$ within the corotation radius.

\begin{figure}
\epsfig{figure=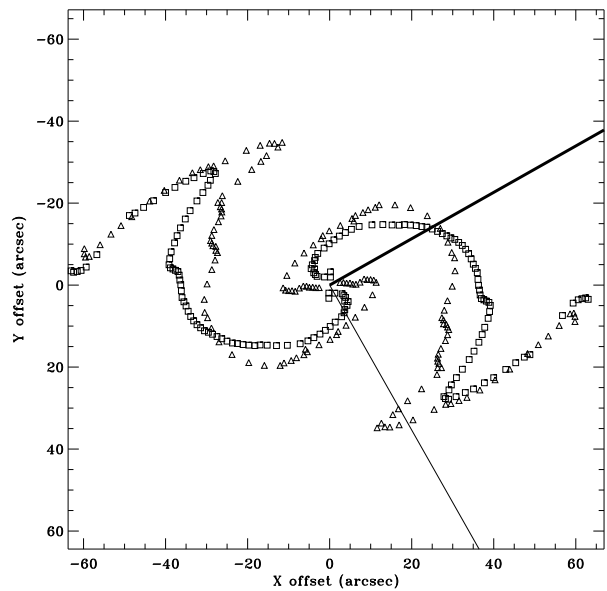,angle=0,width=8.cm,
bbllx=75pt, bblly=135pt, bburx=535pt, bbury=590pt, clip=}
\caption[]{Superposition of the modified third harmonic of the
line-of-sight velocity field (triangles) on the second Fourier harmonic of
the $K$-band brightness map of NGC~157 (squares). The modified third
harmonic has a form of $\cos(2\varphi+\pi/2-F_3)$, where $F_3$ is the
phase of the original third harmonic. Points show the azimuth positions of
the maxima of the relevant harmonic at each radius. A good agreement of
the two phase curves gives strong evidence for the non-circular
velocities, associated with the spiral structure.}
\label{fig16}
\end{figure}

To check the validity of relation (\ref{vst13a}), it is convenient to
compare the locus of maxima of the observed surface density of gaseous
galactic disk and two-armed spiral with the phase value equal to the phase
of the third harmonic shifted by $\pi/2$. In Fig.~\ref{fig16} one can see
the superposition of the maxima of the modified third harmonic of the
line-of-sight velocity field on the maxima of the second Fourier harmonic
of the $K$-band surface brightness map. The modified third harmonic varies
as $\cos(2\varphi+\pi/2-F_3)$, where $F_3(R)$ is the phase of the original
third harmonic. The good correspondence of the phase curve with the
observed position of the spiral arms (see also Fig.~\ref{f-2}) proves that
the observed deviations of the line-of-sight velocity field from the pure
circular motion on one hand and the spiral arms seen in the
brightness map of the galaxy on the other hand are two different
manifestations of the same phenomenon -- the spiral density wave.

It is worth noting that the transformation of the trailing spiral
into the leading one at the periphery of the disk which is seen
in Fig.~\ref{fig16}  may be a real feature of the spiral pattern geometry,
since this peculiarity appears both in the H$\alpha$ and $K$ images
(see Fig.~\ref{f-2}), as well as in the behaviour of the third Fourier
harmonic of the line-of-sight velocity field (Fig.~\ref{fig16}).

\section{Determination of the vector velocity field in the plane
of the galaxy}
\label{sec-6}

It was shown above that the large-scale structure of the velocity field
may be well described in the frame of a model which includes the
first, the second and the third azimuthal Fourier  harmonics. It was also
demonstrated that this result, as well as the structure of the
line-of-sight velocity field, may be naturally explained if one suggests
the residual velocities to be caused by the density wave dominated by a
two--armed mode.  These results allow the restoration of both the
equilibrium rotation curve of the galaxy and the velocity components in
the density wave, by equating correspondent Fourier coefficients in
Eqs.(\ref{eq:vobsout20}) and (\ref{eq:furlast}):
\begin{equation}
\label{eq:AvarphiBrVrot1}
V^{\rm mod2}_{\rm rot} ~+~\frac {C_r \sin F_r
+ C_\varphi \cos F_\varphi  } 2  ~=~ a_1^{\rm obs}  \, ,
\end{equation}
\begin{equation}
\label{eq:AvarphiBrVrot2}
-~ C_r \cos F_r ~+~ C_\varphi \sin F_\varphi ~=~ 2 \, b_1^{\rm obs}  \, ,
\end{equation}
\begin{equation}
\label{eq:AvarphiBrVrot2a}
C_z \cos F_z \,{\rm cot}\, i ~=~  a_2^{\rm obs} \, ,
\end{equation}
\begin{equation}
\label{eq:AvarphiBrVrot2b}
C_z \sin F_z \,{\rm cot}\, i  ~=~ b_2^{\rm obs} \, ,
\end{equation}
\begin{equation}
\label{eq:AvarphiBrVrot3}
-~ C_r \sin F_r ~+~ C_\varphi \cos F_\varphi ~=~ 2\, a_3^{\rm obs}  \, ,
\end{equation} \begin{equation}
\label{eq:AvarphiBrVrot4}
C_r \cos F_r ~+~ C_\varphi \sin F_\varphi ~=~2\, b_3^{\rm obs}  \, ,
\end{equation}

An additional difficulty still exists, since both rotation and motions
in spiral arms contribute to the first cosine harmonic of the
line-of-sight velocity. As a result of this interference, the system of
equations is incomplete.  Indeed, to determine seven
unknown functions on the left--hand side of Eqs.~(\ref{eq:AvarphiBrVrot1})
- (\ref{eq:AvarphiBrVrot4}) we have only six measured Fourier
coefficients.

\begin{figure}
\epsfig{figure=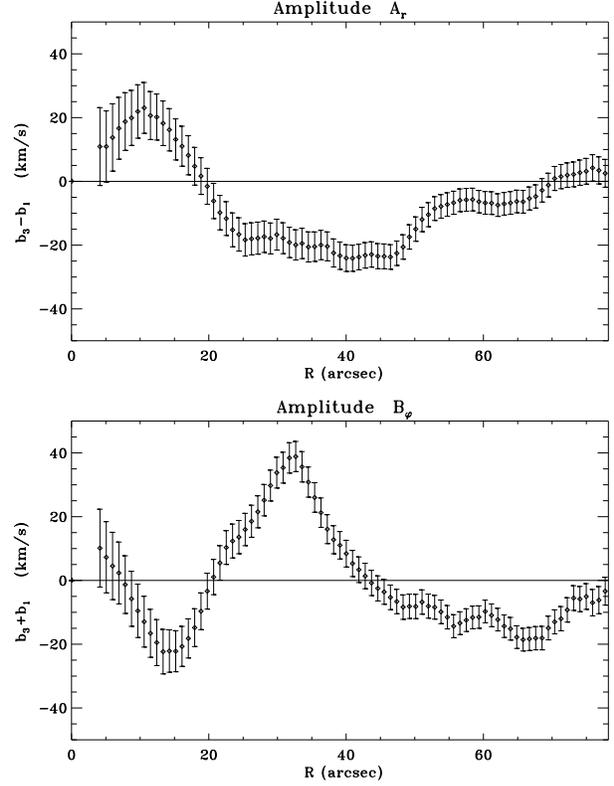,width=8.cm,
bbllx=100pt, bblly=90pt, bburx=550pt, bbury=680pt, clip=}
\caption[]{Radial behaviour of the {amplitudes of the} cosine component for
the radial residual velocity (top) and of the sine component for the
azimuthal residual velocity (bottom).}
\label{f-15}
\end{figure}

From Eqs.~(\ref{eq:AvarphiBrVrot1}) - (\ref{eq:AvarphiBrVrot4})  it
follows that the determination of the coefficients of the
Fourier harmonics of the observed line-of-sight velocity field gives a
possibility to determine the parameters $A_z$ $\equiv$ $C_z \cos
F_z$, $B_z$ $\equiv$ $C_z \sin F_z$, $A_r$ $\equiv$ $C_r \cos F_r$, and
$B_\varphi$ $\equiv$ $C_\varphi \sin F_\varphi$, without any additional
assumptions.  The first pair of these coefficients allows to restore the
vertical velocity. $A_r$ characterises the behaviour of the
perturbed radial velocity along the dynamical major axis of the galaxy,
and $B_\varphi$ characterises the amplitude of the perturbed azimuthal
velocity for the points at the minor axis. The radial dependencies of the
coefficients $A_r (r)$ and $B_\varphi (r)$ are shown in Fig.~\ref{f-15}.

However the full set of unknown parameters: $V^{\rm mod2}_{\rm rot}$,
$C_\varphi$, $F_\varphi$, $C_r$,  and $F_r$ cannot be derived
unambiguously without additional conditions.  To close the set of
equations (\ref{eq:AvarphiBrVrot1}) - (\ref{eq:AvarphiBrVrot4}), we use
two independent approaches proposed in L97 and F97.

\subsection{Method I, based on the relation between the phases
of the radial and azimuthal residual velocities}
\label{sec-6-1}

The first way to solve Eqs.(\ref{eq:AvarphiBrVrot1} --
\ref{eq:AvarphiBrVrot4}) is to use the theoretical relation between the
phases of the radial and tangential residual velocities in a tightly wound
density wave (L97; F97):
\begin{equation}
F_r-F_\varphi~= \left\{ \begin{array}{rl}
\!\!\pi/2 &\!\! {\rm at} ~r<r_c \\
\!\!-\pi/2  & \!\! {\rm at} ~r>r_c
\end{array} \!\! \right. = ~-~{\mathrm sgn}(\hat \omega) \,\pi/2 \, .
\label{vst11}
\end{equation}

By using this relation one can obtain from (\ref{eq:AvarphiBrVrot1} --
\ref{eq:AvarphiBrVrot4}) (L97; F97):
\begin{equation}
\label{eq:model1cc}
V^{\rm mod2}_{\rm rot} ~=~ a^{\rm obs}_1 ~-~ b^{\rm obs}_1 \, a^{\rm obs}_3 / b^{\rm obs}_3\, , \\
\end{equation}
\begin{equation}
\label{eq:model1}
\begin{array}{l}
A_\varphi ~\equiv~ C_\varphi \cos F_\varphi ~=~ a^{\rm obs}_3\,(b^{\rm obs}_1 /
b^{\rm obs}_3~+~1)\,, \\
B_r ~\equiv~ C_r \sin F_r ~=~  a^{\rm obs}_3\,(b^{\rm obs}_1 / b^{\rm obs}_3~-~1)
\,.
\end{array}
\end{equation}

The errors of determination of $b^{\rm obs}_3$ are the most crucial for
application of the above relations. To avoid non-reliable estimates we
assume the following condition for $b^{\rm obs}_3$ to be suitable for
calculations:  $|b^{\rm obs}_3|$ $>$ $3 \cdot \sigma_{b^{\rm obs}_3}$, where
$\sigma_{b^{\rm obs}_3}$ is an error of the measurements.

\begin{figure*}
\epsfig{figure=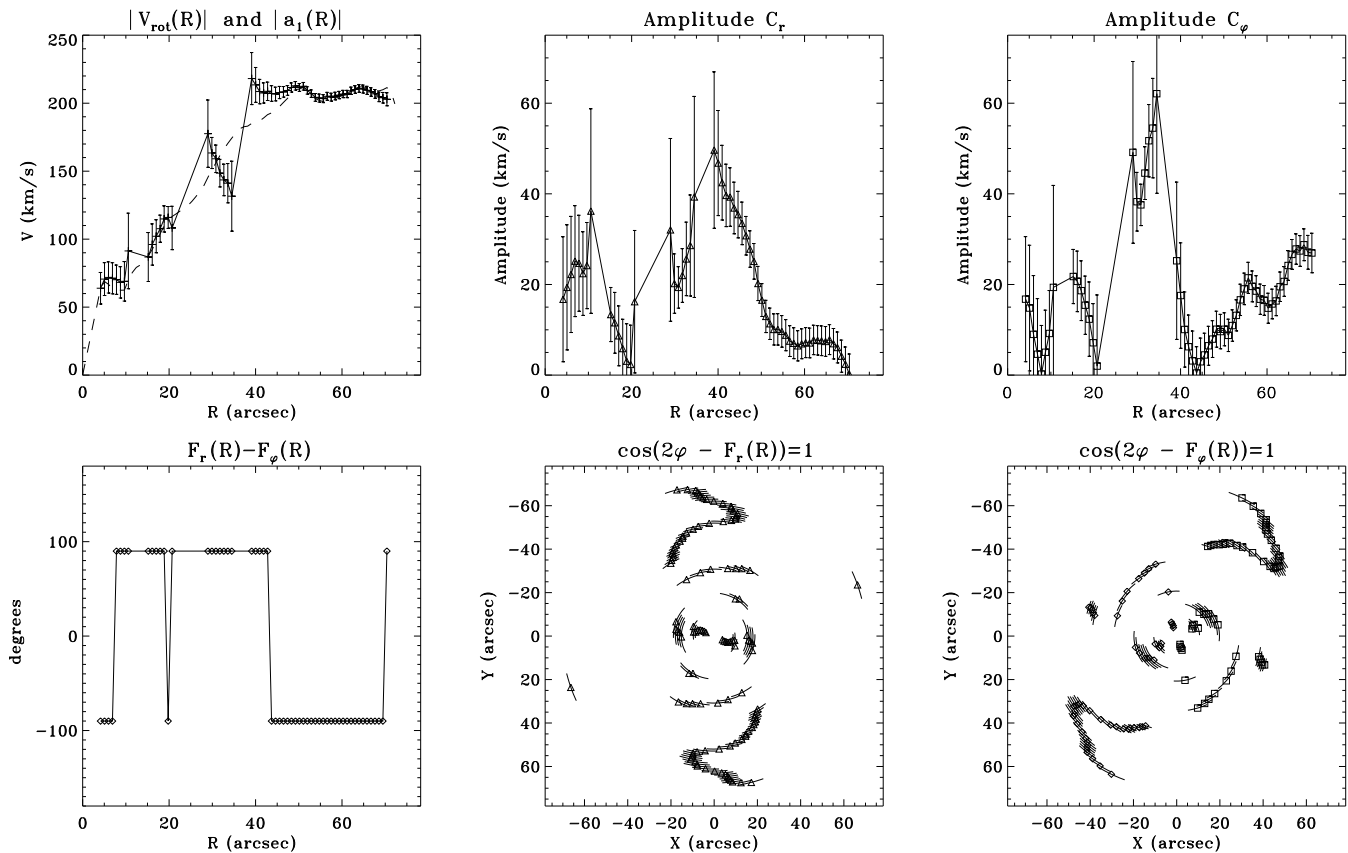,angle=90,width=18cm,
bbllx=55pt, bblly=60pt, bburx=480pt, bbury=730pt, clip=}
\caption[]{The rotation curve $V^{\rm mod2}_{\rm rot}$ (dashed line shows radial
behaviour of $a_1^{\rm obs}$ which corresponds to the rotation velocity in
the model of the pure circular motion, $V^{\rm mod1}_{\rm rot}$), the amplitudes of
the radial and azimuthal residual velocities ($C_r$ and $C_\varphi$), the
phase difference $F_r-F_\varphi$, and the localisation of the maxima of
the residual radial and azimuthal velocities at each radius,
obtained by using the relation between the phases of radial and
tangential residual velocities. The errors are shown at the level of
$3\sigma$.}
\label{f-12}
\end{figure*}

\begin{figure}
\epsfig{figure=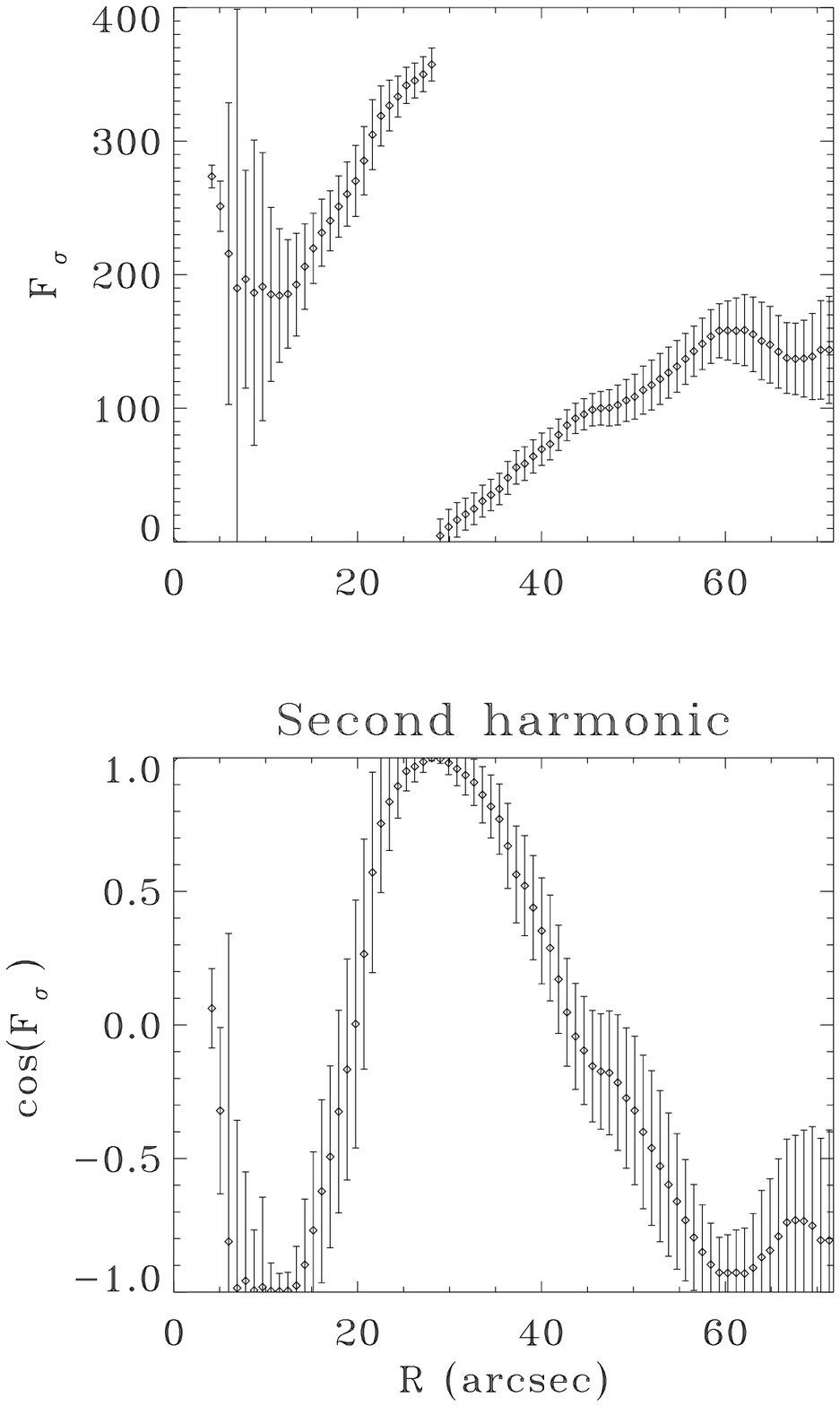,angle=0,width=8.cm,
bbllx=145pt, bblly=150pt, bburx=500pt, bbury=740pt, clip=}
\caption[]{Phase of the second harmonic of the perturbed surface
density (assumed to be equal to the phase of the second harmonic of the K
brightness map) and the value of $\cos F_\sigma$ used to calculate the
parameters of the velocity field according to Method II.}
\label{f-14}
\end{figure}

Fig.~\ref{f-12} shows the rotation velocity  ($V^{\rm mod2}_{\rm rot}$), the
amplitudes and the phases of the radial and tangential residual
velocities calculated from the above equations in the regions where
$|b^{\rm obs}_3|$ $>$ $3 \cdot \sigma_{b^{\rm obs}_3}$. For comparison,
the rotation curve $V_{\rm rot}^{\rm mod1}$ obtained in the model of
pure circular motion (it corresponds to the Fourier coefficient $a_1^{\rm
obs}$) is also plotted, as a dashed line.  One can see from this figure
that the maxima of the radial and tangential residual velocities
outline  the shape of trailing spirals, whose  pitch angle is
close to that of the visible spiral arms. Within $7\arcsec < r <
43\arcsec$, the phase difference between the radial and tangential
velocities $F_r-F_\varphi$ is equal to $90^\circ$, and outside $r =
43\arcsec$ $F_r-F_\varphi$ $=$ $-90^\circ$. Therefore, this method enables
us to localize the corotation radius in the region around $r=
43\arcsec$ (see Eq.\,\ref{vst11}).  The mean amplitudes of the perturbed
radial and tangential velocities are 20 -- 30~km s$^{-1}$. The
rotation velocity found here differs from the rotation velocity determined
in the model of pure circular motion by amounts comparable to the
velocity amplitude in the spiral arms.

\subsection{Method II, based on the relation between the phases
of the residual radial velocity and perturbed surface density}
\label{sec-6-2}

Another way to close the system of Eqs.(\ref{eq:AvarphiBrVrot1} --
\ref{eq:AvarphiBrVrot4}) is to use the relation between the phases of
the perturbed surface density and the radial residual
velocity, in the approximation of tightly wound spirals (L97; F97):
\begin{equation}
F_\sigma-F_r~= \left\{
\begin{array}{ll}
\!\! \pi & \!\!{\rm at} ~r<r_c \\
\!\! 0  & \!\!{\rm at} ~r>r_c
\end{array}
\!\! \right. = ~[1-{\mathrm sgn}(\hat \omega)] \,\pi/2 \, .
\label{vst8}
\end{equation}

The determination of the perturbed density phase using the image of a
galaxy is a separate problem. Here we assume that the locations of the
maxima of the perturbed surface density at a given radius coincide with
the locations of the maxima of the surface brightness related to the
spiral arms. This assumption allows to determine the behaviour of the
phase of the second Fourier harmonic of the perturbed surface density
$F_\sigma$ by applying the Fourier analysis to the brightness distribution
in a galaxy. The values of $F_\sigma$ and $\cos F_\sigma$ obtained in this
way from the $K$-band image of NGC~157 (see Fig.~\ref{f-2}) are shown in
Fig.~\ref{f-14}.

From (\ref{eq:AvarphiBrVrot1}) - (\ref{eq:AvarphiBrVrot4}) and (\ref{vst8})
we have (L97; F97):
\begin{equation}
V^{\rm mod2}_{\rm rot} ~=~ a^{\rm obs}_1 ~-~ a^{\rm obs}_3 ~-~ (b^{\rm
obs}_3-b^{\rm obs}_1) \hbox{tan} F_\sigma \, ,
\end{equation}
\begin{equation}
A_\varphi ~=~ 2 a^{\rm obs}_3 ~+~ (b^{\rm obs}_3-b^{\rm obs}_1) \,
\hbox{tan} F_\sigma \, ,
\end{equation}
\begin{equation}
\label{eq:model2bb}
B_r ~=~ (b^{\rm obs}_3 ~-~ b^{\rm obs}_1) \, {\rm tan}\, F_\sigma \, .
\end{equation}

\begin{figure*}
\epsfig{figure=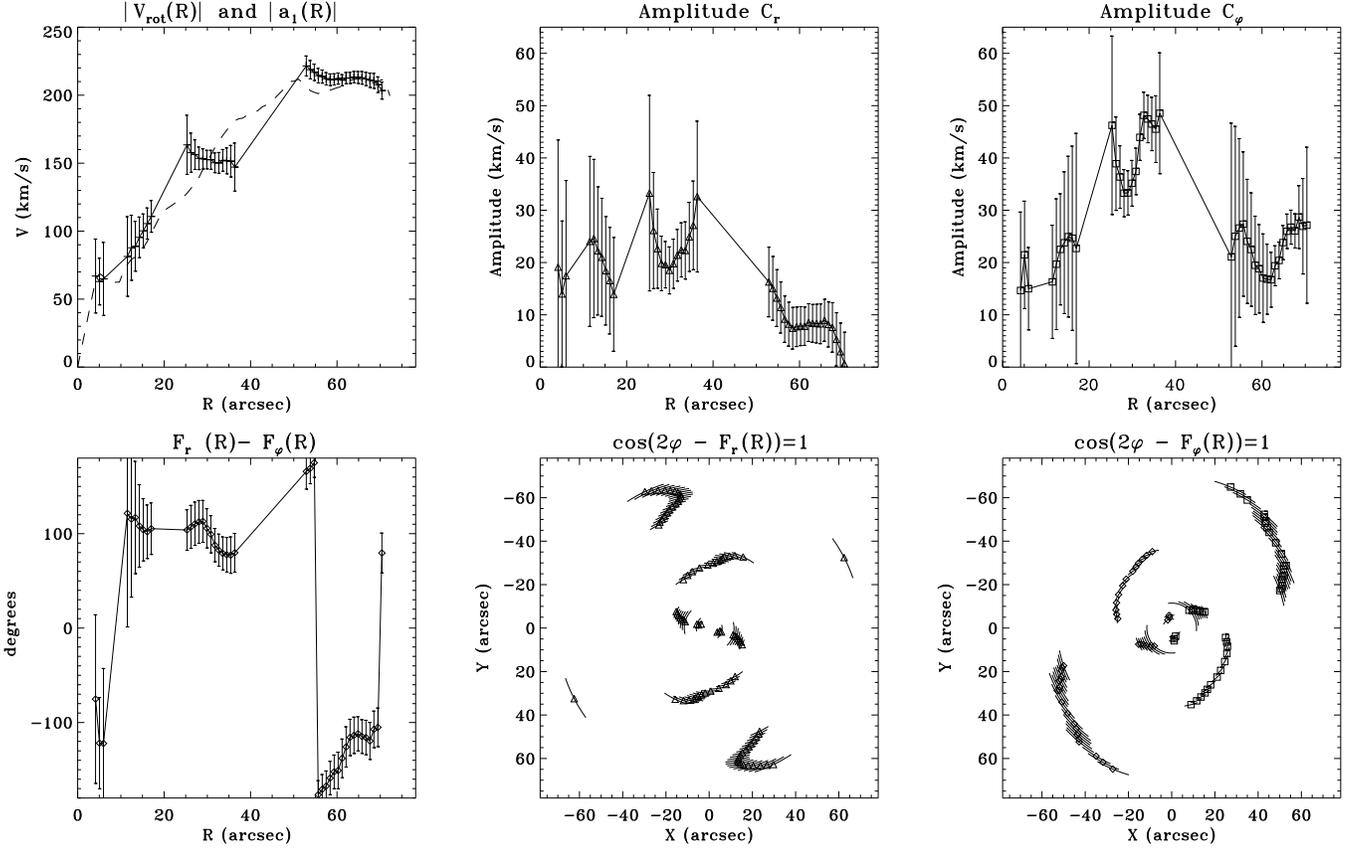,angle=90,width=18cm,
bbllx=55pt, bblly=60pt, bburx=480pt, bbury=730pt, clip=}
\caption[]{AS Fig.~16, now obtained by using the relation between
phases of the radial residual velocity and the perturbed surface density.
The errors are shown at the level of $3\sigma$.}
\label{f-13}
\end{figure*}

Fig.~\ref{f-13} presents the radial dependence of the same
parameters as in Fig.~\ref{f-12}, but obtained by  Method II for those
regions where $\hbox{cos} F_\sigma$ can be determined most reliably, that
is, where $|\hbox{cos} F_\sigma|$ $>$ $3\cdot \sigma_{\hbox{cos}
F_\sigma}$ (see Fig.~\ref{f-14}).  As one can see from Fig.~\ref{f-13},
the lines of the maxima of the radial and tangential velocities look like
trailing spirals whose appearance agrees rather well with the phases
obtained earlier by  Method I (see Fig.~\ref{f-12}).  The change of the
phase difference between the radial and tangential velocities from
$F_r-F_\varphi$ $\simeq$ $90^\circ$ to $F_r-F_\varphi$ $\simeq$
$-90^\circ$ occurs in the radial range $36\arcsec < r < 63\arcsec$.
Therefore, the corotation radius obtained by this method is located within
this range -- in agreement with the estimate found above by Method I. The
characteristic amplitudes of the radial and tangential velocities are of
the order of 20 -- 30~km s$^{-1}$, again in good agreement with the
previous estimates.

\subsection{Final model of the velocity field}

\begin{figure}
\epsfig{figure=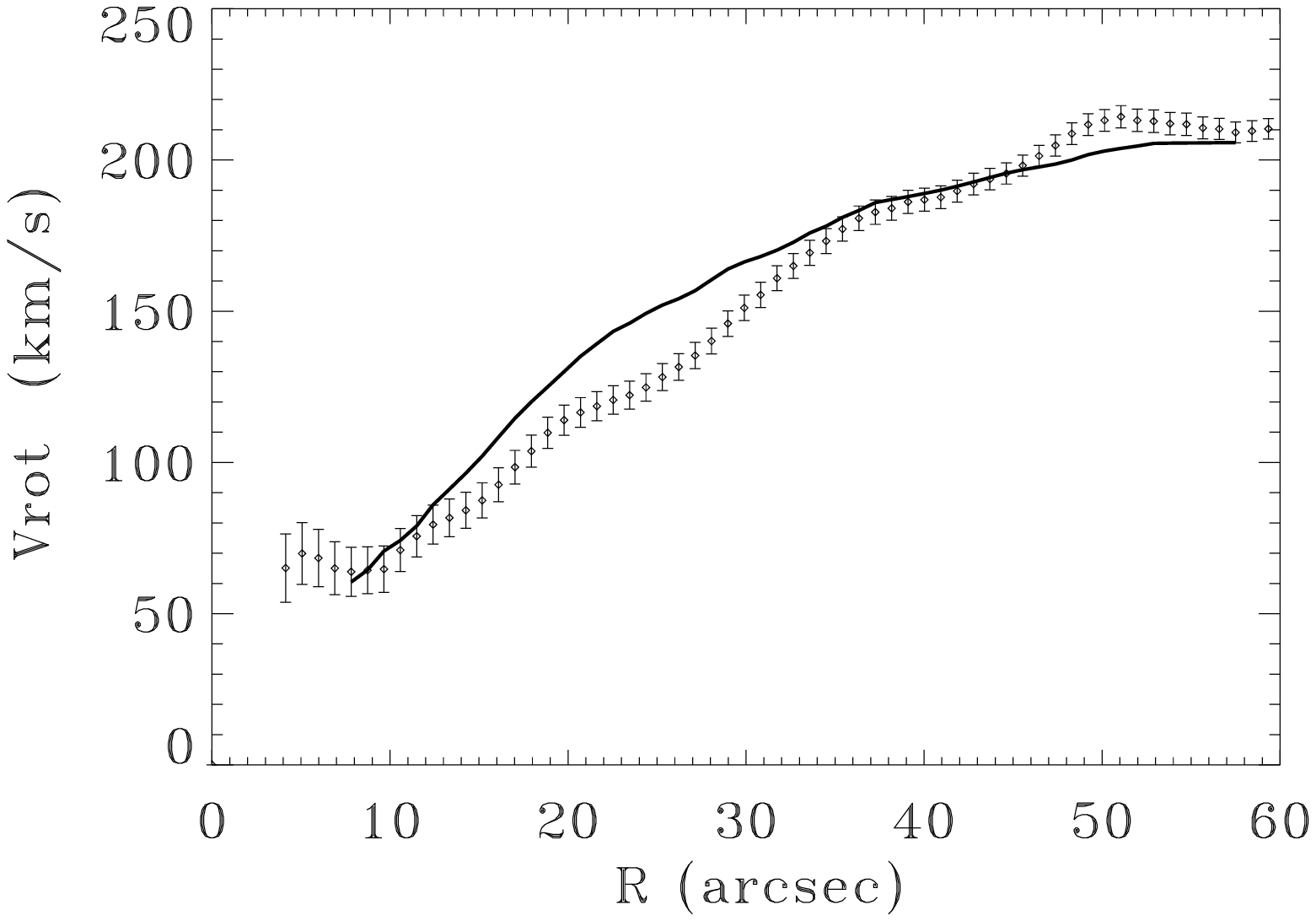,angle=0,width=8.6cm,
bbllx=85pt, bblly=370pt, bburx=540pt, bbury=690pt, clip=}
\caption[]{Rotation curve of NGC 157 in the best fit model (solid
line) and in the model of pure rotation (dotted  line with error bars
$3\sigma$). }
\label{fig11}
\end{figure}

The most significant systematic error in the estimates obtained  above
arises from the assumption of a sharp switch of the phase relations
near corotation. In reality, the area of switch cannot be too
narrow. Our estimates show (L97) that the switch area
size is of order of $1/k_r$, which for NGC~157 corresponds to
10\arcsec -- 20\arcsec.
Therefore, the velocity field parameters derived above have to be
smoothed near corotation by using a filter with a  width of about
$1/k_r$.

Another argument in favour of the necessity of smoothing is the
behaviour of the rotation curve near the corotation found by
Methods I and II. Its non-monotonic features, seen in Figs.~\ref{f-12} and
\ref{f-13}, directly result from the assumed sharp switch of the phases,
since the radial distribution of the azimuthally-averaged stellar disk
brightness, which follows the surface density distribution, is rather
smooth and hence gives no reason to expect any peculiarities of the
rotation curve near $r=30\arcsec - 50\arcsec$.

So we claim the necessity to smooth the calculated rotation curve
near the corotation. That, in turn, leads to a broadening of the
switch area where the phase differences of the radial and tangential
residual velocities and the perturbed surface density change.

To obtain the best model of the velocity field for the galaxy, the radial
dependence of the residual velocities was varied within the range of
uncertainties given by the comparison of Methods~I and II. The
"best fit" model was chosen as a model which gives a smooth
shape of the rotation curve, which can be expected for a smooth
mass distribution in the galaxy. The rotation curve for the best fit model
is shown in Fig.~\ref{fig11}.  For comparison, the rotation curve obtained
in the traditional way, using a model of pure rotation, is also
presented.  In the latter case, the presence of wave-shaped features
is seen.  Similar wave-like details of rotation curves were found in many
galaxies, and they can naturally be explained in the frame of the density
wave theory (Lin {\it et al.} 1969; Yuan 1969).

\begin{figure*}
\epsfig{figure=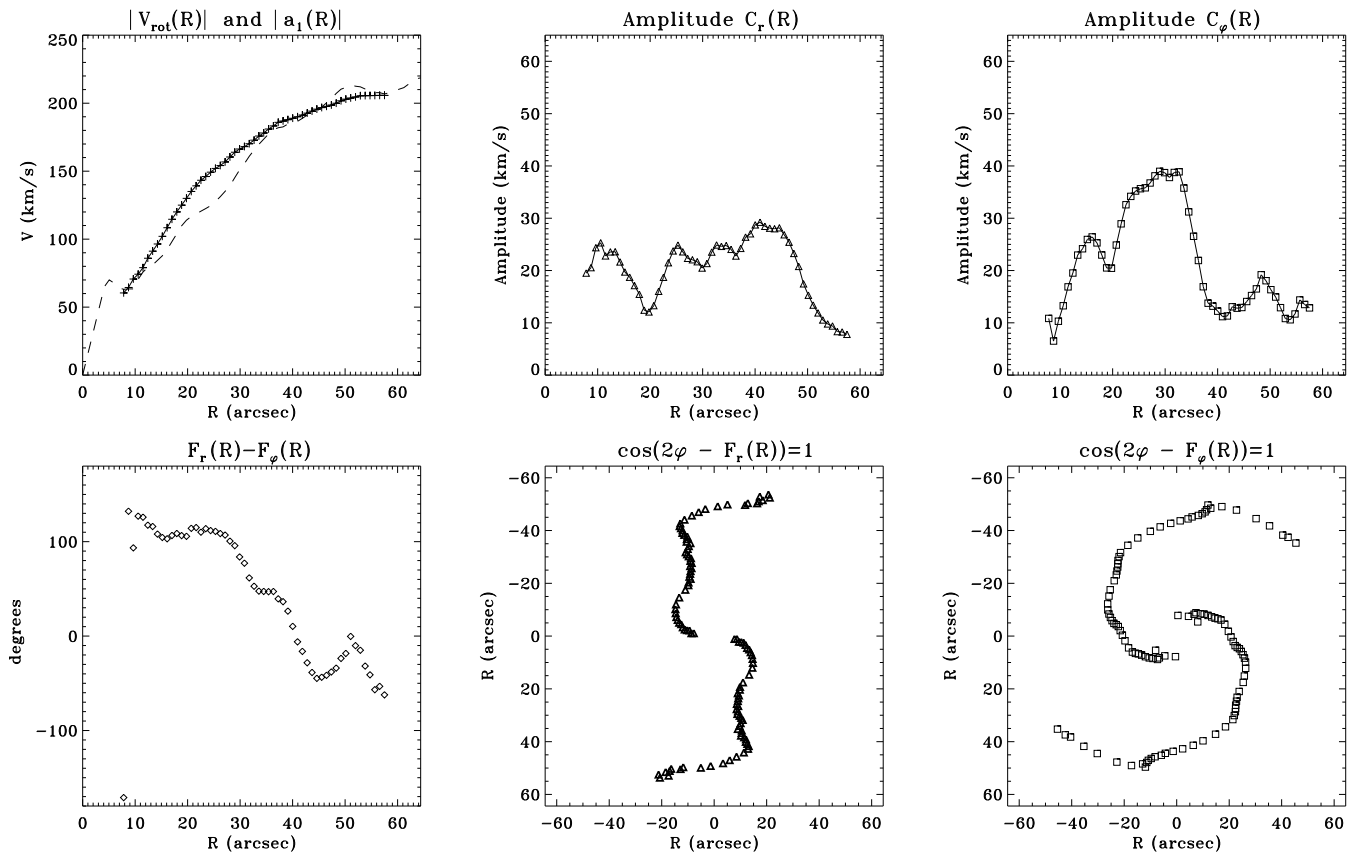,angle=90,width=18cm,
bbllx=55pt, bblly=60pt, bburx=480pt, bbury=730pt, clip=}
\caption[]{The rotation curve $V^{\rm mod2}_{\rm rot}$ (dashed line shows the radial
behaviour of $a_1^{\rm obs}$, which corresponds to the rotation velocity in
the model of pure circular motion, $V^{\rm mod1}_{\rm rot}$), the amplitudes, the
phase difference $F_r-F_\varphi$, and the positions of the maxima of
radial and tangential residual velocities at different radii obtained in
the best-fit model (see the text).}
\label{f-17}
\end{figure*}

The estimation of the rotation velocity $V^{\rm mod2}_{\rm rot}$ at different radii
enables us to determine unambiguously the other characteristics of the
galactic velocity field from the set of equations
(\ref{eq:AvarphiBrVrot1} -- \ref{eq:AvarphiBrVrot4}).  Fig.~\ref{f-17}
shows the behaviour of the rotation curve, as well as that of the
amplitudes and phases of radial and tangential velocities in the density
wave, as obtained for the best-fit model.

\section{Corotation radius of the spiral pattern}

The principal aim of the present work is discovery of new galactic
structures -- giant anticyclones -- near the corotation.
To make these visible,  knowledge of the corotation radius of
the density wave is needed. Below we applied the methods proposed in
L97 and F97 to determine the
position of corotation from the observed line-of-sight velocity
field of the ionized gas in the galaxy.

\subsection {Method 1, using the relation between the phases of
radial and tangential residual velocities}

From Eqs.\,(\ref{eq:AvarphiBrVrot1}) - (\ref{eq:AvarphiBrVrot4}) and
(\ref{vst11}) the following relation can be obtained (L97; F97):
\begin{equation}
\label{eq:s19aaa}
\frac {b_3^{\rm obs} ~-~ b_1^{\rm obs}} {b_1^{\rm obs} ~+~ b_3^{\rm obs}} ~=~
\frac{(b_3^{\rm obs})^2 ~-~ (b_1^{\rm obs})^2} {(b_1^{\rm obs} ~+~ b_3^{\rm obs})^2 } ~=~
\mp~ \frac{C_\varphi}{C_r} \,.
\end{equation}
Hereafter, the upper sign in plus-minus and minus plus signs corresponds
to the region inside the corotation, and the lower sign -- outside it.
Taking into account that the amplitudes $C_r$ and $C_\varphi$ have always
positive values, it follows from equation (\ref{eq:s19aaa}) that
\begin{equation} \label{eq:b1b3} \begin{array}{ll} |b_3^{\rm
obs}(r)|-|b_1^{\rm obs}(r)| \le 0, & \hbox{for~~ } r < r_c \,, \\
|b_3^{\rm obs}(r)|-|b_1^{\rm obs}(r)| \ge 0, & \hbox{for~~ } r > r_c \, .
\end{array}
\end{equation}

These inequalities offer a possibility to determine the location of
corotation from the observational data. According to Eq.(\ref{eq:b1b3}),
corotation is located in the region where the difference of the
amplitudes of the third sine and the first sine harmonics ($|b_3^{\rm
obs}(r)| - |b_1^{\rm obs}(r)|$) changes its sign from minus to plus.

\begin{figure}
\epsfig{figure=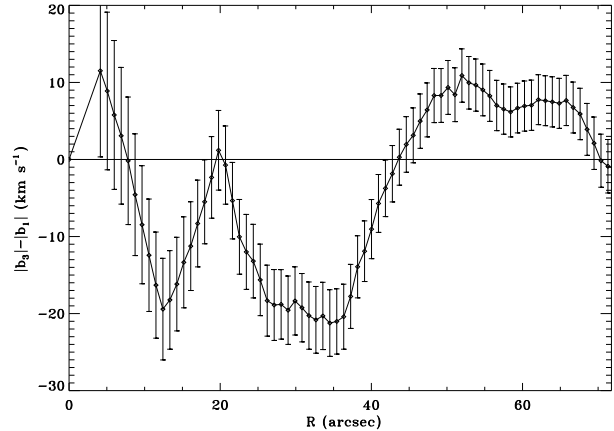,width=8.cm,
bbllx=70pt, bblly=365pt, bburx=540pt, bbury=705pt, clip=}
\caption[]{ Trend of $|b_3^{\rm obs}(r)|-|b_1^{\rm obs}(r)|$ with
galactocentric radius $r$ in NGC~157. Under the approximation of tightly
wound spirals this difference should be negative inside corotation and
positive outside it.  According to the data presented, the corotation
radius is about $42\arcsec \pm 5\arcsec$.  Error bars correspond to
$3\sigma$ level.  The jumps of the velocity components in the regions $r <
20\arcsec$ and $r> 65\arcsec$ unveil some peculiarities of the galactic
structure in these regions. In the central area it may be ascribed to the
presence of a bar, at the periphery these features are located where
we observe a strong kink of spiral arms.}
\label{f-18}
\end{figure}

Fig.~\ref{f-18} shows the radial behaviour of $|b_3^{\rm obs}(r)| -
|b_1^{\rm obs}(r)|$ in NGC~157.  As one can see from the figure,  this
function is negative within the errors in the inner part of the galaxy and
positive in the outer region, in accordance with the expectations. From
these data it follows that the corotation radius is about $42\arcsec \pm
5\arcsec$.

\subsection{Method 2, using the relation between the phases
of the radial residual velocity and the perturbed surface density.}

System (\ref{eq:AvarphiBrVrot1}) - (\ref{eq:AvarphiBrVrot4}) closed by
relation (\ref{vst8}) gives (L97; F97):
\begin{equation}
\label{eq:s18bbb}
(b_3^{\rm obs} ~-~ b_1^{\rm obs})\,\cos F_\sigma ~=~\mp~ C_r \, \cos^2 F_\sigma
\, ,
\end{equation}
from which follows {bf that}
\begin{equation}
\label{eq:new2}
\begin{array}{ll}
(b_3^{\rm obs}(r) - b_1^{\rm obs}(r))\, \cos F_\sigma (r) ~\le~ 0, &
~~ \hbox{for~} r < r_c, \\
(b_3^{\rm obs}(r)-b_1^{\rm obs}(r)) \, \cos F_\sigma (r) ~\ge~ 0, &
~~ \hbox{for~} r > r_c.
\end{array}
\end{equation}
These two inequalities allow to find the location of corotation
on the base of the brightness map and the line-of-sight velocity field.

\begin{figure}
\epsfig{figure=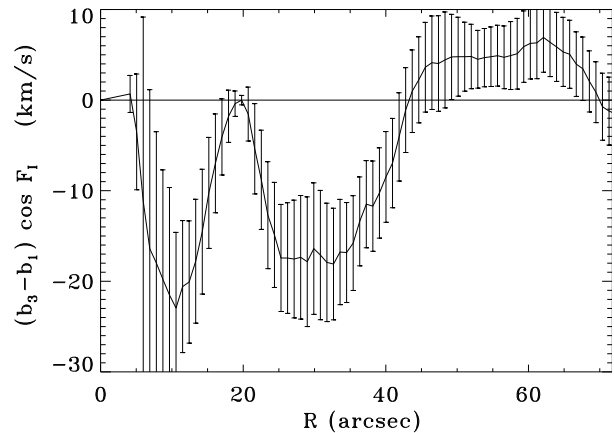,angle=0,width=8.cm,
bbllx=80pt, bblly=370pt, bburx=530pt, bbury=695pt, clip=}
\caption[]{Variation of $(b_3^{\rm obs}-b_1^{\rm obs}) \cos F_\sigma $ with
galactocentric radius $r$ in NGC~157.  In the WKB-approximation the
difference should be negative inside corotation and positive outside
it. Error bars correspond to $3\sigma$ level. These data show the
corotation radius to be about $42\arcsec \pm 6\arcsec$.}
\label{fig13}
\end{figure}

Fig.~\ref{fig13} shows the dependence of $(b_3^{\rm obs}-b_1^{\rm obs}) \cos
F_\sigma $ on galactocentric radius $r$.  The estimation of $r_c$ by
this method gives $42\arcsec \pm 6\arcsec$.

It should be noted that the methods described above may possess not only
random errors, but also some systematic errors caused by the
violation of the approximation of tightly wound spirals. However they use
observational data related to different parameters of the spiral
structure, and their systematic errors must be different. So the results
of the application of these methods give evidence that the radius of
corotation is within 36\arcsec--48\arcsec and that the characteristic
value of the systematic error introduced by the tightly wound
approximation hardly exceeds 6\arcsec. For the second method described
above, the error may be slightly higher because in this case an additional
assumption about one-to-one correspondence between the azimuthal positions
of the maxima of perturbed density and perturbed luminosity was
made. Note that a slightly higher value for the corotation radius,
$r=50\arcsec$, was recently obtained from the numerical simulation of
NGC~157 by Sempere and Rozas (1997).

\begin{figure}
\epsfig{figure=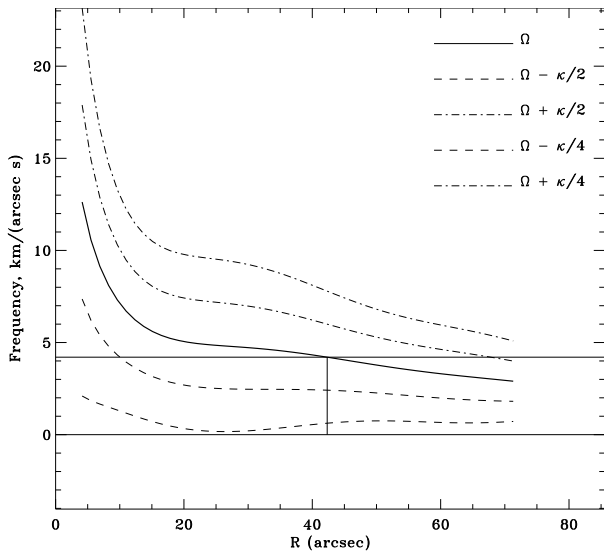,angle=0,width=8.cm,
bbllx=80pt, bblly=205pt, bburx=520pt, bbury=605pt, clip=}
\caption[]{Positions of the main resonances of the spiral structure in
NGC~157 for the corotation position $r_c=42\arcsec$. }
\label{res}
\end{figure}

Assuming the location of the corotation radius to be within the radius
range of 36\arcsec--48\arcsec, and taking the rotation curve as obtained
in the best fit model, the inner Lindblad resonance (ILR) is either
complitely absent in this galaxy, or located near the very galactic
center, where the shape of the rotation curve is poorly determined.  The
outer Lindblad resonance (OLR) is located near the edge of the ionized-gas
disk, in the region where the well-defined spiral arms end.
Therefore, the bright spiral arms are constrained to exist inside the OLR.
This result agrees with the results of the numerical simulations  of the
spiral pattern in NGC~157 (Sempere \& Rozas 1997).  On the other hand, a
strong kink of the spiral arms (change from trailing to leading spirals)
is located near the outer 4:1 resonance (compare Fig.~\ref{f-14} and
Fig.~\ref{res}).

The estimate of the difference between the phases of the radial
perturbed velocity and the azimuthal perturbed velocity  in the best
fit model allows to find the corotation radius more accurately by using
the relation between the phases of radial and azimuthal velocity
perturbations at the corotation. As was shown in L97, the phase difference
between the radial and azimuthal perturbed velocities becomes zero at
corotation. In our case, this corresponds to $r_c = 42\arcsec$.
This value will be used in the next section to restore the velocity field
of the gas in the reference frame rotating with the spiral pattern.

It is worth noting that the corotation radius was found above by using
some particular values of the disk inclination angle $i$ and the position
angle of the line of nodes PA$_0$, which give the minimum dispersion
in the model taking into account the first, second, and third harmonics
of the line-of-sight velocity field. Any changes of these parameters,
even within the errors of their determinations, may affect the
determination of the location of corotation. So we have carried out a
special analysis of the influence of the variations of inclination and
position angle on the estimate of the corotation radius. We find
that variations of the inclination by $\pm 5^\circ$ do not change (within
the accuracy of 1\arcsec-- 2\arcsec) the position of the corotation
radius.  However the value of $r_c$ strongly depends on the accepted
$PA_0$.  Variations of $PA_0$ by $\pm 2^\circ$ result in corotation
radius variations of about $\sim 5\arcsec$. However, the qualitative
picture of the restored velocity field remains the same even in this case.

\section{Vortex structures}

\begin{figure*}
    \epsfig{figure=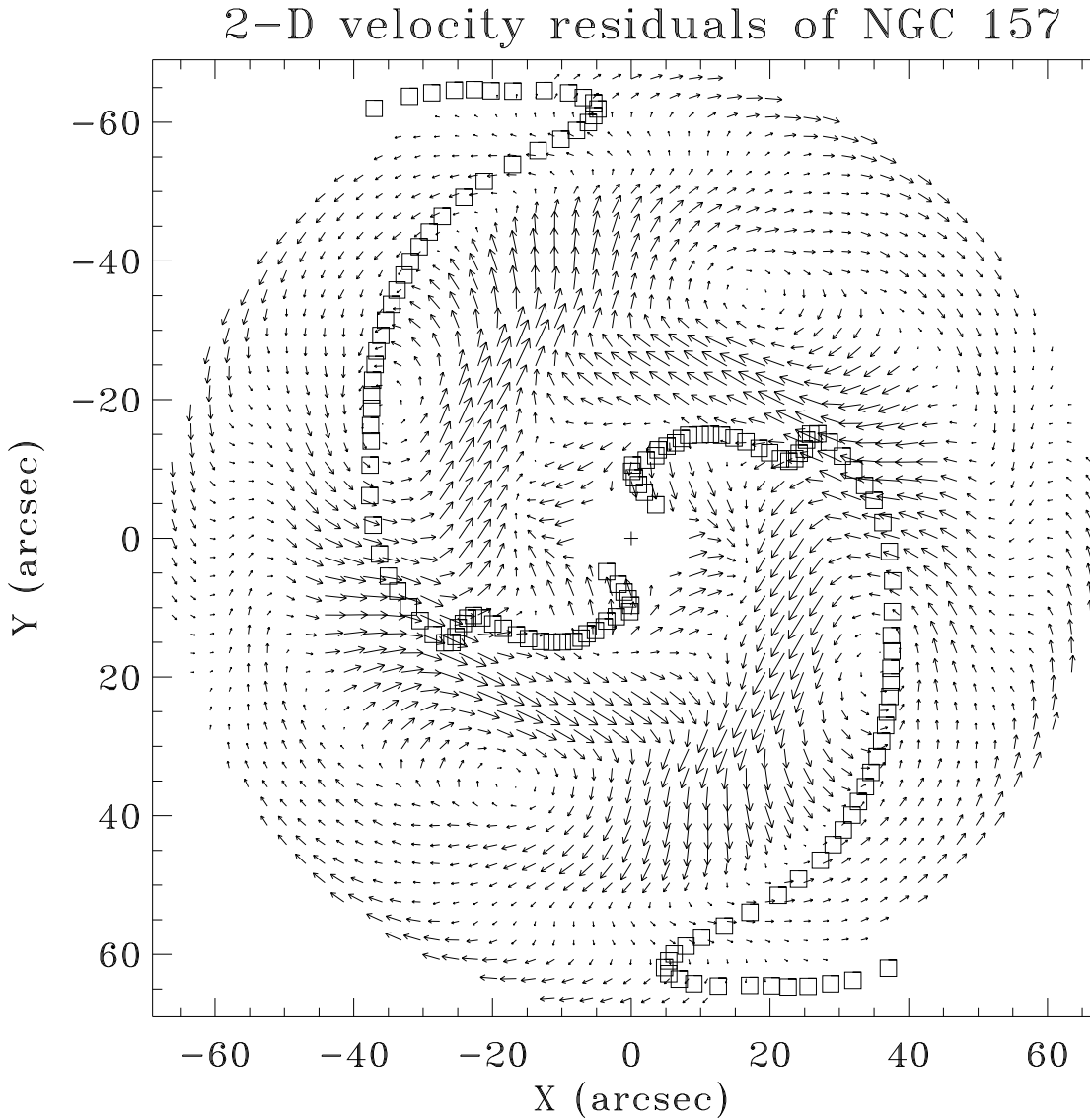,width=15cm,
bbllx=65pt, bblly=185pt, bburx=495pt, bbury=725pt, clip=}
\caption[]{Restored vector field of the residual velocity in gaseous
disk of NGC~157. Overlaid squares for each
galactocentric radius show azimuthal positions of maximum value of
the second Fourier harmonics of the H$\alpha$ brightness map of the
galaxy.}
\label{f-19}
\end{figure*}

As a  first step to visualising the restored velocity field of
NGC~157, in  Fig.~\ref{f-19} we show a vector field of the
residual velocities of the gas in the plane of the galactic disk,
calculated for the best fit model. We can see four vortices near
$r_c$ -- two cyclones and two anticyclones. This result is found for
all values of parameters used  for the restoration within the range of
uncertainties.

\begin{figure*}
    \epsfig{figure=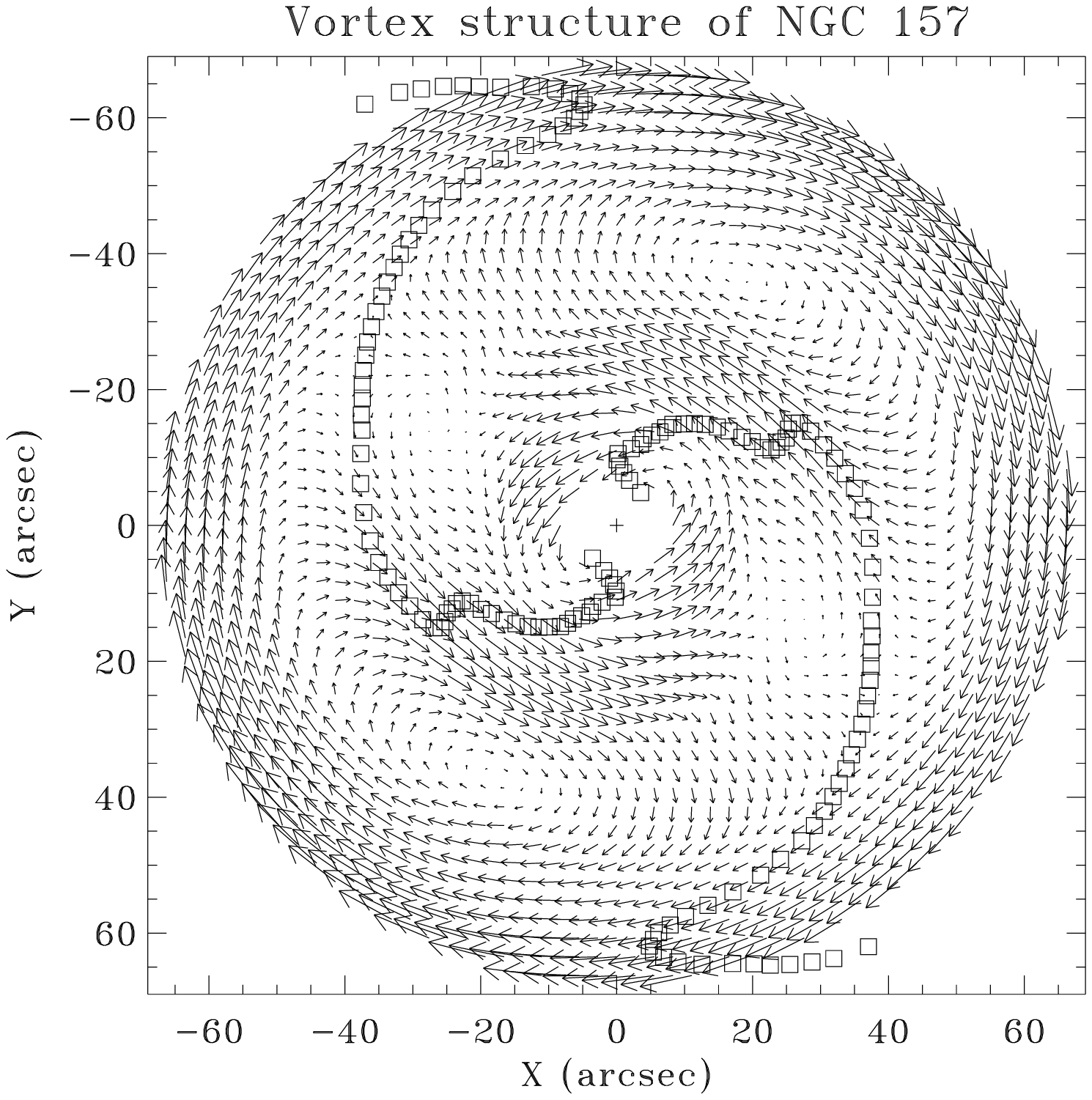,width=15cm,
bbllx=65pt, bblly=185pt, bburx=495pt, bbury=725pt, clip=}
\caption[]{Restored velocity field of NGC~157 in the reference frame
rotating with the pattern speed. Overlaid squares for each
galactocentric radius show azimuthal positions of maximum value of
the second Fourier harmonics of the H$\alpha$ brightness map of the
galaxy.}
\label{f-20} \end{figure*}

The restored velocity field in the reference frame  rotating with the
pattern speed  of the density wave, obtained in the best fit model, is
shown in Fig.~\ref{f-20}. The gas velocities inside the spiral arms
are directed preferably along the arms:  toward the center inside the
corotation circle and outward outside the corotation.  It is clearly seen
that the velocity field  of the galaxy demonstrates two anticyclones
located in the corotation region between the spiral arms.  Such
anticyclonic vortices had been predicted on the basis of laboratory
experiments on shallow water (Nezlin et al. 1986). The velocity amplitude
of the vortices is about 30 km s$^{-1}$. The two cyclones seen
in the residual velocity field are absent in the full velocity
field in the reference frame corotating with the spiral arms. To make the
cyclones visible, the radial gradient of the perturbed azimuthal velocity
($\partial \tilde V_\varphi/ \partial r$) should be larger than the radial
gradient of the rotation velocity (${\rm d}(V_{\rm rot} - \Omega_c r)/{\rm
d}r$) in the same region of the disk.  It is necessary to transform the
anticyclonic shear, caused by the disk rotation, into a cyclonic shear. In
the gaseous disk of NGC~157 this condition is not fulfilled.  The
cyclones, if they exist, must be confined to a small number of resolution
elements in our data so their detection is very difficult.  In galaxies
with a stronger gradient of $\tilde V_\varphi$ cyclones can be
observed\footnote{Within the three years of the present paper publication
the prediction was confirmed (Fridman et al. 1999; Fridman et al. 2001a;
Fridman et al. 2001b).}.

As shown by numerical simulations (Baev et al. 1987; Baev \& Fridman
1988), spirals and vortices are generated  simultaneously by one and the
same instability, already at the linear stage. The levels of
saturation of the instability are different for different perturbed
functions -- the densities of gaseous and stellar disks and their
velocities. The amplitudes of the velocities and stellar disk
density perturbations stopped growing at a linear stage, while the
amplitude of the gaseous disk density perturbation becomes nonlinear. As
has been shown many times during the past 15 years, the streamlines of
liquid particles in a self--consistent field of hydrodynamical and/or
self-gravitational forces form cyclones and antycyclones (Nezlin et al.
1986; Baev \& Fridman 1988; Afanasiev \& Fridman 1993; Lyakhovich et
al. 1996; F97).  The field of the average velocity of the stellar disk
under such conditions has not been calculated so far. In a general
case, the structure of the field of the average velocity in the stellar
systems differs substantially from the geometry of individual
trajectories of stars.  Contopoulos (1978) was the first to calculate
individual trajectories of stars in an external gravitational
potential of spiral arms.  He showed that every star participates
simultaneously in two motions:  high-frequency (around the center of
the "epicycle") and low-frequency (around the Lagrangian
points L$_2$ and L$_4$).

As a consequence of the specific choice of the corotation radius (see
the end of the previous Section), the phase difference
between the azimuthal and radial residual velocities becomes zero
at corotation, and both velocities simultaneously drop to zero at the
centers of the vortices directly on the corotation circle. But, as was
noted earlier, the qualitative picture of the restored velocity field does
not depend on the particular choice of parameters within the range of
their uncertainties. In all cases two anticyclones appear between the
spiral arms, with their centers located near the corotation radius.

It should be noted that in reality  the general picture of gas
velocities in a galaxy includes not only the motions in the plane of a
disk, but also vertical motions which contribute to the second Fourier
harmonic of the line-of-sight velocity field
(Eqs.\,\ref{eq:aAbB3},~\ref{eq:aAbB4}), and hence can be determined in the
frame of the model described above
(Eqs.\,\ref{eq:AvarphiBrVrot2a},~\ref{eq:AvarphiBrVrot2b}).  The most
convenient way to study vertical motions is to measure gas velocities in
galaxies seen nearly face-on, where velocity components in the plane of
the disk give a small contribution to the line-of-sight velocity. As an
example, we refer to the analysis of the velocity field (both from H{\sc i}
and H$\alpha$ observations) of the spiral galaxy NGC~3631, where vertical
gas motions related to the spiral arms were really found (Fridman et al.
1998, 2000).

Comparison of the location of the spiral arms with that of
the anticyclones shows that the centers of the anticyclones lie
between the spiral arms. Thus, the self-gravitational forces in the gas
dominate over the forces of hydrodynamic pressure in this galaxy
(Lyakhovich et al. 1996). Therefore the origin of the spiral-vortex
structure of NGC~157 must be explained in the frame of the gravitation
concept of the  density waves.

An analysis similar to the one  undertaken in the present work is
possible only for galaxies where the numerous estimates of the observed
gas velocities are well distributed all over the disk and spiral density
waves are present. It demonstrates a great potential concerning the
investigation of Fourier components of the line-of-sight velocity
azimuthal distributions when a sufficient amount of observational data
is available.

\section{Conclusions}

\begin{itemize}

\item
{The line-of-sight velocity field obtained for the spiral galaxy NGC~157
from interferometric observations in the H$\alpha$ emission line,
allowed us to obtain reliable estimates of the parameters of the disk
orientation, of the coordinates of the dynamical center and of
the rotation curve in the approximation of pure circular motion.}

\item
{Deviations of the observed line-of-sight velocities
from pure circular motion have a systematic character and can be
described by the first three harmonics of the Fourier expansion of
the azimuthal distribution of velocities.}

\item
{When perturbed gas motions are taken into account, these do
not substantially change the estimates of the disk inclination, the
dynamical center position, and the major axis orientation.}

\item
{The good correspondence between  two phase curves -- that of
the modified third harmonic of the line-of-sight velocity field and that
of the second Fourier harmonic of the perturbed surface density of
NGC~157, gives evidence for the wave nature of the spiral structure in
this galaxy.}

\item
{The predominance of the first three Fourier harmonics in the
line-of-sight velocity field is naturally explained by the prevalence of
perturbations with mode $m$ $=$2 in the density wave. This circumstance
opens the way to restore the full vector velocity field of the ionized gas.}

\item
{Two methods of vector velocity field restoration in a galaxy with a
two-armed spiral are applied to the line-of-sight velocity field of
NGC~157. They give qualitatively and quantitatively similar results.
Typical values of perturbed velocities in the density wave are 20--30 km
s$^{-1}$. The maximum amplitude of the radial velocity in the density wave
is $\simeq 30$ km s$^{-1}$, that of the azimuthal velocity is $\simeq 40$
km s$^{-1}$.}

\item
{Using the approximation, that  the pitch angle of the
spiral pattern is small, we propose two methods to determine the
position of the corotation radius. The first method is based solely
on the Fourier analysis of the observed velocity field, and gives
$r_c =42\arcsec \pm 5\arcsec$. The second method involves also the
additional data on radial variations of the perturbed surface density
phase. This method gives $r_c=42\arcsec \pm 6\arcsec$, which shows
that these two independent approaches are in good agreement
with each other.}

\item
{The determination of the corotation radius of the spiral pattern and
the restoration of the gas vector velocity field in the galaxy have
enabled us to detect anticyclonic structures, the existence of which had
been earlier predicted analitically and from laboratory experiments on
shallow water (Nezlin et al. 1986). For the galaxy NGC~157 we found
that the anticyclone centers are located between the spiral arms; this
gives evidence for the dominance of self-gravitation over hydrodynamic
pressure forces near corotation (Lyakhovich et al.  1996).}

\end{itemize}

\begin{acknowledgements}
We would like to express many thanks to Preben Grosb{\o}l
and Johan Knapen for their hard editorial work, which has
substantially improved our original text. We thank George Countopoulos
and Panos Patsis for many fruitful discussions. This work was performed
under partial financial support from RFBR grant N 99-02-18432, grant
``Leading Scientific Schools''  N 00-15-96528, and the grants
``Fundamental Space Researches.  Astronomy''  N 1.2.3.1, N 1.7.4.3.
\end{acknowledgements}

\appendix

\setcounter {equation}{0}
\def\theequation{A\arabic{equation}}

\section{Derivation of basic relations}

This Appendix contains a short derivation of the basic relations used in
the Paper to close the system (\ref{eq:AvarphiBrVrot1}) -
(\ref{eq:AvarphiBrVrot4}), in order to restore the vector velociy
field of the gas, and to determine the corotation radius of the spiral
structure.  Whereas the final results were obtained earlier (L97; F97) the
method of derivation presented  here is new and more concise.

Following the traditional way of  representating small amplitude
perturbations in the disk (see e.g. Lin \& Shu 1964; Fridman \&
Polyachenko 1984) we shall derive this relation below for the case of the
tightly wound spiral approximation. The perturbed functions for a grand
design spiral galaxy can be presented in the following form:
\[
\sigma(r,\,\varphi,\,t)~=~\sigma_0(r)+\sigma_1(r,\,\varphi,\,t)~=~
\]
\begin{equation}
~~~~~=~\sigma_0(r) ~+~ \tilde \sigma\, \exp {\rm i}
[\int \!\! k(r){\mathrm d}r+2\varphi-\omega t - F_\sigma] \, ,
\label{vst1a}
\end{equation}
\[
V_r(r,\,\varphi,\,t) = V_{r1}(r,\,\varphi,\,t) =
\]
\begin{equation}
~~~~~=~\tilde V_r \exp {\rm i} [\int \!\! k(r){\mathrm d}r+2\varphi-\omega
t - F_r] \, ,
\label{vst2a}
\end{equation}
\[
V_\varphi(r,\,\varphi,\,t)~=~V_{\varphi 0}(r)+V_{\varphi
1}(r,\,\varphi,\,t) ~=~
\]
\begin{equation}
~~~~~=~ V_{\varphi 0}(r) ~+~ \tilde V_\varphi\, \exp {\rm i}[\int \!\!
k(r){\mathrm d} r+2\varphi-\omega t - F_\varphi] \, ,
\label{vst3}
\end{equation}
where $k(r)$ is a local radial wave number, and
\begin{equation}
\int\! \! k(r){\mathrm d}r ~\gg~ 2 \,.
\label{vst4}
\end{equation}
The latter inequality is equivalent to the tightly wound arms
approximation, which is in agreement with the form of the arms of
NGC~157.

Unlike the expressions (\ref{vst2}) - (\ref{eq:sprvz2}), here we use the
complex representation of the perturbed values and assume that the
amplitudes marked by a tilde in
Eqs.~(\ref{vst1a}) - (\ref{vst3}) slightly
depend on $r$.

Substituting (\ref{vst1a}) - (\ref{vst3}) into the linearized continuity
equation for the surface density\footnote{ The continuity equation for
the surface density, which is derived by integrating over $z$ the initial
3D continuity equation, contains the residual velocities $\vec V_1$ $=$
$\sigma_0^{-1}\int \rho_0 \vec v_1 dz$, which are residual velocities
averaged over $z$.}
\begin{equation}
\frac {\partial
\sigma_1}{\partial t} + \frac {1}{r}\, \frac {\partial }{\partial r}
(r\sigma_0 V_{r1}) + \frac {_1}{r} \frac {\partial }{\partial
\varphi}(\sigma_0 V_{\varphi 1}+\sigma_1 V_{\varphi 0}) = 0 \label{vst5}
\end{equation}
and using (\ref{vst4}), we obtain
\begin{equation}
\hat \omega \tilde \sigma \exp(-i F_\sigma)~=~\sigma_0 k_r \tilde V_{r}
\exp (-i F_r) \, ,
\label{vst6}
\end{equation}
where
\begin{equation}
\hat \omega ~\equiv~ \omega -2 \Omega_0(r) \, .
\label{vst7}
\end{equation}

In most galaxies $\Omega_0$ is a decreasing function of $r$. Thus
according to (\ref{vst7}) one obtains: $\hat \omega$ $<$ 0 inside
corotation radius (at $r$ $<$ $r_c$) and $\hat \omega$ $>$ 0 outside
corotation radius (at $r$ $>$ $r_c$). Then for a trailing spiral, $k_r$
$>$ 0, we find from (\ref{vst6}) the following relation between phases:
\begin{equation}
F_\sigma-F_r~= \left\{ \begin{array}{ll}
\!\! \pi & \!\!{\rm at} ~r<r_c \\
\!\! 0  & \!\!{\rm at} ~r>r_c
\end{array} \!\! \right. = ~[1-{\mathrm sgn}(\hat \omega)] \,\pi/2 \, .
\label{vst8aa}
\end{equation}

From the linearized equation of motion in the disk plane under the
condition (\ref{vst4}) we obtain\footnote{For barotropic perturbations the
function $V_{\varphi0}$ depends on $r$ only (the Poincare theorem).
Therefore the initial equation of motion can be integrated over $z$,
to obtain the equation for the residual velocity $\vec V$ averaged over
$z$ (L97).} (Fridman \& Khoruzhii, 1999):
\begin{equation}
\frac {V_{r1}} {V_{\varphi1}} ~=~ \frac {\tilde V_{r} \, \exp
(-{\rm i}F_r)} {\tilde V_{\varphi} \,\exp(-{\rm i}F_\varphi)} ~=~ {\rm
i} \frac {2\Omega_0 \hat \omega} {\kappa^2} \,,
\label{vst9}
\end{equation}
whence:
\begin{equation}
\exp[-{\rm i}(F_r-F_\varphi)] ~=~
\frac {2\Omega_0 |\hat \omega|}{\kappa^2} \, \frac {\tilde V_\varphi}
 {\tilde V_r} \, \exp[{\rm i} \,{\rm sgn}(\hat \omega) \,\pi/2] \,  .
\label{vst10}
\end{equation}
Here $\kappa$ is an epicyclic frequency, $\kappa^2$
$\equiv$ $2 \Omega_0( 2\Omega_0+r\Omega_0')$. As follows from
Fig.~\ref{f-3}, $\kappa^2$ $>$ 0, therefore from (\ref{vst10}) we find
\begin{equation}
F_r-F_\varphi~= \left\{ \begin{array}{rl}
\!\!\pi/2 &\!\! {\rm at} ~r<r_c \\
\!\!-\pi/2  & \!\! {\rm at} ~r>r_c
\end{array} \!\! \right. = ~-~{\mathrm sgn}(\hat \omega) \,\pi/2 \, .
\label{vst11aa}
\end{equation}

Adding (\ref{vst8aa}) to (\ref{vst11aa}) and multiplying by -1 we arrive
at the following relations:
\begin{equation}
F_\varphi-F_\sigma~=\left\{
\begin{array} {ll}
\!\!\pi/2 & \!\!{\rm at} ~r<r_c \\ \!\!\pi/2  & \!\!{\rm at} ~r>r_c
\end{array}\!\!  \right. = ~\pi/2 \, .
\label{vst11a} \end{equation}

On the other hand, from Eqs.~(\ref{eq:AvarphiBrVrot3}),
(\ref{eq:AvarphiBrVrot4}) taking into account (\ref{vst11aa}), one can
derive
\begin{equation}
\label{vst12}
{\rm tan}\, F_\varphi = \frac  { b_3^{\rm obs}} {a_3^{\rm obs}}  , ~\hbox{sgn}
\left(\cos F_\varphi\right) = \hbox{sgn} \left[\frac
{a_3^{\rm obs}} {(C_\varphi \mp C_r}) \right]\,.
\end{equation}

Comparing the latter with the expressions derived from (\ref{vst14}),
(\ref{vst15}):
\begin{equation}
\label{vst12zz}
{\rm tan}\, F_3 = \frac
{ b_3^{\rm obs}} {a_3^{\rm obs}}  , ~\hbox{sgn} \left(\cos F_3\right) =
\hbox{sgn} \left({a_3^{\rm obs}} \right)\,,
\end{equation}
we see that outside corotation the phase of the azimuthal perturbed
velocity ($F_\varphi$) coincides with the phase of the third Fourier
harmonic ($F_3$).

Inside corotation the relation between phases of the perturbed
azimuthal velocity and the third Fourier harmonic of the observed
line-of-sight velocity depends on the ratio of amplitudes
$C_\varphi$ and $C_r$.

For a two--armed tightly wound spiral the ratio of
amplitudes of azimuthal and radial residual velocities is (see
Eq.~(\ref{vst9})):
\begin{equation}
\label{vst12aa}
\frac{C_\varphi}{C_r} ~=~ \frac{|\tilde V_\varphi|}{|\tilde V_r|} ~=~
\frac{\kappa^2}{2\Omega|\hat \omega|}~ =~
\frac{(2\Omega+r\Omega')}{2|\Omega-\Omega_{ph}|} \,.
\end{equation}
Near corotation, this ratio is high because the denominator is small.
The analysis of the rotation curve of NGC~157 (Fig.~\ref{f-3}) shows
that in the inner region of the disk the value $r{\rm d}\Omega/{\rm d}r$
is small,  therefore \begin{equation} \label{vst12ab} C_\varphi / C_r
~\simeq~ \Omega/|\Omega-\Omega_{ph}| ~>~ 1\,.  \end{equation}

Consequently, in most of the galactic disk the
phase of the residual azimuthal velocity will be approximately equal to
the phase of the third Fourier harmonic of the observed line-of-sight
velocity:
\begin{equation}
\label{vst13}
F_\varphi~=~F_3 \, .
\end{equation}

Combining the relations (\ref{vst11a}) and (\ref{vst13}) we finally obtain the
connection between the phases of the two observable values --- perturbed
surface density and third Fourier harmonic of the line-of-sight velocity:
\begin{equation}
\label{vst13aaa}
F_3~\approx~F_\sigma ~+~ \pi/2 \, .
\end{equation}
This relation should be valid if the grand design spiral has
a wave nature.

The derived relations between the phases of the
different parameters of the perturbations can be used to close
(\ref{eq:AvarphiBrVrot1}) -- (\ref{eq:AvarphiBrVrot4}) and determine the
characteristics of the full vector velocity field of the galaxy.

The system of equations (\ref{eq:AvarphiBrVrot1}) --
(\ref{eq:AvarphiBrVrot4}) can be rewritten as follows:  \begin{equation}
\label{eq:s16}
A_z~=~a_2^{\rm obs} {\rm cot}\, i,
\end{equation}
\begin{equation}
\label{eq:s17}
B_z~=~b_2^{\rm obs} {\rm cot}\, i,
\end{equation}
\begin{equation}
\label{eq:s18}
A_r~=~-b_1^{\rm obs}~+~b_3^{\rm obs},
\end{equation}
\begin{equation}
\label{eq:s19}
B_\varphi~=~b_1^{\rm obs}~+~b_3^{\rm obs},
\end{equation}
\begin{equation}
\label{eq:s20}
V^{\rm mod2}_{\rm rot}+A_\varphi~=~a_1^{\rm obs}~+~a_3^{\rm obs},
\end{equation}
\begin{equation}
\label{eq:s21}
V^{\rm mod2}_{\rm rot}+B_r~=~a_1^{\rm obs}~-~a_3^{\rm obs},
\end{equation}
where we have inserted the sine and cosine of the residual velocity
components related to their amplitudes and phases:
\begin{equation}
\label{eq:AA}
A_i ~=~ C_i\, \cos F_i \,,
\end{equation}
\begin{equation}
\label{eq:BB}
B_i ~=~  C_i \, \sin F_i \, .
\end{equation}
The former parameter characterises the residual velocity
variations along the dynamical major axis of the galaxy, and the latter
describes variations along the minor axis.

It follows from (\ref{eq:s16}) - (\ref{eq:s19}) that by using only
observational data, without any additional conditions, one can
determine the vertical velocities in the galaxy together with the
components $A_r$ and $B_\varphi$, which describe the amplitude of
the perturbed radial velocity at the major axis of the galaxy and the
amplitude of the perturbed azimuthal velocity at the minor axis,
respectively. The vertical motions are described by the second harmonic
and differ from $a_2^{\rm obs} \cos (2\varphi) + b_2^{\rm obs} \sin (2
\varphi)$ only by a factor of $\hbox{cot}\,i$ which is approximately equal
to 1.3 for NGC~157.

Two remaining equations, (\ref{eq:s20}) and (\ref{eq:s21}), contain three
unknowns, $V^{\rm mod2}_{\rm rot}$, $A_\varphi$, and $B_r$, which cannot be
determined without additional conditions.

The first way to solve Eqs.(\ref{eq:s16}) -- (\ref{eq:s21}) is to use
the relation (\ref{vst11aa}) between the phases of the radial and
tangential residual velocities. By using this condition, from
(\ref{eq:s18}), (\ref{eq:s19}), (\ref{eq:AA}) and (\ref{eq:BB}) we obtain
\begin{equation}
\label{eq:model1aa}
\pm~ \frac {C_r}{C_\varphi} ~=~ \frac {b^{\rm obs}_1 ~-~ b^{\rm obs}_3}
{b^{\rm obs}_1 ~-~ b^{\rm obs}_3} \, ,
\end{equation}
and from (\ref{eq:s20}), (\ref{eq:s21})
\begin{equation}
\label{eq:model1bb}
\pm~ \frac {C_r}{C_\varphi} ~=~ \frac {a^{\rm obs}_1 ~-~ a^{\rm obs}_3 ~-~
V^{\rm mod2}_{\rm rot}} {a^{\rm obs}_1 ~+~ a^{\rm obs}_3 ~-~ V^{\rm mod2}_{\rm rot}} \, .
\end{equation}
The upper sign corresponds to the region inside corotation, and the lower
sign -- outside it.

By equalising the right hand terms of (\ref{eq:model1aa}) and
(\ref{eq:model1bb}) we find at $r$ $<$ $r_c$, as well as at $r$ $>$ $r_c$:
\begin{equation}
\label{eq:model1cccc}
V^{\rm mod2}_{\rm rot} ~=~ a^{\rm obs}_1 ~-~ b^{\rm obs}_1 \, a^{\rm obs}_3 / b^{\rm obs}_3\, , \\
\end{equation}
and finally from (\ref{eq:s20}) and (\ref{eq:s21}) we get
\begin{equation}
\label{eq:model1zz}
\begin{array}{l}
A_\varphi ~=~ a^{\rm obs}_3\,(b^{\rm obs}_1 / b^{\rm obs}_3~+~1)\,, \\
B_r ~=~  a^{\rm obs}_3\,(b^{\rm obs}_1 / b^{\rm obs}_3~-~1) \,.
\end{array}
\end{equation}

Another way to close the system of Eqs.(\ref{eq:s16}) -- (\ref{eq:s21})
is to use the relation (\ref{vst8aa}) between phases of the perturbed surface
density and the radial residual velocity. The phase of the perturbed density
can be determined on the basis of analysis of an  image of a
galaxy, assuming that the locations of the maxima of the
perturbed surface density at a given radius coincide with the locations of
the maxima of the surface brightness related to the spiral arms.

From (\ref{eq:s18}) and (\ref{eq:AA})  we have
\begin{equation}
\label{eq:model2aa}
C_r \, \cos\, F_r ~=~ b^{\rm obs}_3 ~-~ b^{\rm obs}_1  \, ,
\end{equation}
and immediately the value $B_r$ is calculated, using its definition
(\ref{eq:BB}) and the phase relation  (\ref{vst8aa}):
\begin{equation}
\label{eq:model2bbbb}
B_r ~=~ C_r \, \sin\, F_r ~=~ (b^{\rm obs}_3 ~-~ b^{\rm obs}_1) \, {\rm tan}\,
F_\sigma \, .
\end{equation}

Substituting (\ref{eq:model2bbbb}) into (\ref{eq:s21}), we find
for both $r$ $<$ $r_c$ and $r$ $>$ $r_c$:
\begin{equation}
V^{\rm mod2}_{\rm rot} ~=~ a^{\rm obs}_1 ~-~ a^{\rm obs}_3 ~-~ (b^{\rm obs}_3-b^{\rm obs}_1)
\hbox{tan} F_\sigma \, ,
\end{equation}
whence
\begin{equation}
A_\varphi ~=~ 2 a^{\rm obs}_3 ~+~ (b^{\rm obs}_3-b^{\rm obs}_1) \, \hbox{tan} F_\sigma
\, .
\end{equation}

In a similar way, the relations for determination of the corotation radius
from the observational data can be derived.

From (\ref{eq:model1aa}), we have
\begin{equation}
\label{eq:s19aaaaa}
\frac {b_3^{\rm obs} ~-~ b_1^{\rm obs}} {b_1^{\rm obs} ~+~ b_3^{\rm obs}} ~=~
\frac{(b_3^{\rm obs})^2 ~-~ (b_1^{\rm obs})^2} {(b_1^{\rm obs} ~+~ b_3^{\rm obs})^2 } ~=~
\mp~ \frac{C_\varphi}{C_r} \,.
\end{equation}
Taking into account that the amplitudes $C_r$ and $C_\varphi$ always have
positive values, equation (\ref{eq:s19aaaaa}) gives
\begin{equation}
\begin{array}{ll}
|b_3^{\rm obs}(r)|-|b_1^{\rm obs}(r)| \le 0, & \hbox{for~~ } r < r_c \,, \\
|b_3^{\rm obs}(r)|-|b_1^{\rm obs}(r)| \ge 0, & \hbox{for~~ } r > r_c \, .
\end{array}
\end{equation}
These inequalities offer a possibility to determine the location of
corotation from the radial behaviour of $|b_1^{\rm obs}|$ and
$|b_3^{\rm obs}|$ in the observed line-of-sight velocity
field.

On the other hand, from (\ref{eq:model2aa}) we have
\begin{equation}
\label{eq:s18bb}
C_r \, \cos F_r ~=~ \mp~ C_r \, \cos F_\sigma ~=~ b_3^{\rm obs} ~-~
b_1^{\rm obs} \, .
\end{equation}

Multiplying the last equality of Eq.(\ref{eq:s18bb}) by $\cos F_\sigma$, we
find the equation
\begin{equation}
(b_3^{\rm obs} ~-~ b_1^{\rm obs})\,\cos F_\sigma ~=~\mp~ C_r \, \cos^2
F_\sigma \, ,
\end{equation}
from which follows
\begin{equation}
\begin{array}{ll}
(b_3^{\rm obs}(r) - b_1^{\rm obs}(r))\, \cos F_\sigma (r) ~\le~ 0, &
~~ \hbox{for~} r < r_c, \\
(b_3^{\rm obs}(r)-b_1^{\rm obs}(r)) \, \cos F_\sigma (r) ~\ge~ 0, &
~~ \hbox{for~} r > r_c.
\end{array}
\end{equation}
These two inequalities allow to find the location of corotation
on the basis of the surface brightness map and the line-of-sight velocity
field.

\end{document}